\documentclass[iop]{emulateapj}
\usepackage{graphicx}
\usepackage{amsmath}
\usepackage{ifthen}
\usepackage{afterpage}

\slugcomment{(Draft \today)}

\gdef\gappr{\hbox{$_>\atop{^\sim}$}}
\gdef\lappr{\hbox{$_<\atop{^\sim}$}}

\gdef\Sec{${}^{\prime\prime}$\llap{.}}

\def\etal{\hbox{et al.}}

\gdef\ltsima{$\scriptscriptstyle \; \buildrel < \over \sim \;$}
\gdef\simlt{\lower.3ex\hbox{\ltsima}}

\gdef\gtsima{$\scriptscriptstyle \; \buildrel > \over \sim \;$}
\gdef\simgt{\lower.3ex\hbox{\gtsima}}

\gdef\about{\raise.3ex\hbox{$\scriptscriptstyle \sim $}}

\shortauthors{Kelson \etal\ }
\shorttitle{CSI Redshift Survey: I. Description and Methodology}

\begin{document}

\footnotetext[1]{This paper includes data gathered with the 6.5 meter Magellan Telescopes 
located at Las Campanas Observatory, Chile.}

\title{The Carnegie-Spitzer-IMACS Redshift Survey of Galaxy Evolution since $\lowercase{z}=1.5$:
\break
I. Description and Methodology\footnotemark[1]
}

\footnotetext[2]{Visiting Astronomer, Kitt Peak National
Observatory, National Optical Astronomy Observatory, which is operated
by the Association of Universities for Research in Astronomy (AURA)
under cooperative agreement with the National Science Foundation.}

\author{
Daniel D. Kelson\footnotemark[2],
Rik J. Williams\footnotemark[2],
Alan Dressler\footnotemark[2],
Patrick J. McCarthy\footnotemark[2],
Stephen A. Shectman,
John S. Mulchaey,
Edward V. Villanueva,
Jeffrey D. Crane, \&
Ryan F. Quadri
}
\affil{The Observatories of the Carnegie Institution for
Science, 813 Santa Barbara St., Pasadena, CA 91101}

\begin{abstract}
We describe the Carnegie-Spitzer-IMACS (CSI) Survey, a wide-field, near-IR selected spectrophotometric redshift 
survey with the Inamori Magellan Areal Camera and Spectrograph (IMACS) on Magellan-Baade. By defining a flux-limited sample 
of galaxies in Spitzer \emph{IRAC} $3.6\mu$m imaging of SWIRE fields, the CSI Survey efficiently traces the stellar mass of 
average galaxies to $z\sim 1.5$. This first paper provides an overview of the survey selection, observations,
processing of the photometry and spectrophotometry.
We also describe the processing of the data: new methods of fitting synthetic templates of spectral energy distributions
are used to derive redshifts, stellar masses, emission line luminosities, and coarse information on recent star-formation. Our
unique methodology for analyzing low-dispersion spectra taken with multilayer prisms 
in \emph{IMACS}, combined with panchromatic photometry from the ultraviolet to the IR, has
yielded high quality redshifts for 43,347 galaxies in our first 5.3 degs$^2$ of the SWIRE XMM-LSS field.
We use three different approaches to estimate our redshift errors and find robust agreement. 
Over the full range of $3.6\mu$m fluxes of our selection, we find
typical redshift uncertainties of $\sigma_z/(1+z)
\la 0.015$. In comparisons with previously published spectroscopic redshifts we find scatters of
$\sigma_z/(1+z) = 0.011$ for galaxies at $0.7\le z\le 0.9$, and
$\sigma_z/(1+z) = 0.014$ for galaxies at $0.9\le z\le 1.2$.
For galaxies brighter and fainter than $i=23$ mag, 
we find $\sigma_z/(1+z) = 0.008$ and $\sigma_z/(1+z) = 0.022$, respectively.
Notably, our low-dispersion spectroscopy and analysis yields
comparable redshift uncertainties and success rates for both red and blue galaxies, largely
eliminating color-based systematics that can seriously bias observed dependencies of galaxy evolution on
environment.

\end{abstract}



\section{Introduction}
\label{sec:intro}

\noindent Understanding the evolution  of galaxies and large scale structure remains a fundamental challenge 
in astrophysics.  Many ambitious galaxy surveys have been carried out to address this problem, but limited time 
on large telescopes results in a classic problem: sky coverage, depth, and spectral resolution -- choose two.  For 
example, the very-wide-area Sloan Digital Sky Survey (SDSS)  provides a wealth of spectral information for 
galaxies over a cosmologically significant volume, but its modest depth limits the SDSS to the relatively local, 
`modern' universe.  Conversely, the Hubble Ultra-Deep Field photometric survey probes deep into cosmic 
time --- to the epoch of reionization, but only over a tiny volume whose small population of galaxies leads to large 
uncertainties on their physical properties. 
While the union of such surveys has begun to paint a coherent picture 
of galaxy growth, significant patches of blank canvas limit our ability to fully describe
how environmental processes, the infall of 
gas that fuels star formation, galaxy mergers and acquisitions, and feedback, 
all shape the evolution of galaxies over cosmic time.

The Carnegie-Spitzer-IMACS Survey (CSI) has been designed to address one of the most dramatic and least
understood features of galaxy evolution --- the remarkably rapid decline in cosmic star formation since
$z\sim 1.5$.  It is during this extended epoch of galaxy maturation that galaxy groups and clusters have also
emerged as a conspicuous feature of the landscape.  The CSI Survey is uniquely able to link together the 
evolution of individual galaxies with these features of large-scale structure growth. 

In our ambitious spectrophotometric redshift survey of distant galaxies, we strike a balance between
the aforementioned three factors: (1) high completeness to moderate redshift ($z\simlt 1.5$), 
(2) spectral resolution intermediate between conventional photometric and spectroscopic 
surveys (combining the efficiency of imaging surveys with a spectral resolution high enough to 
resolve large-scale structure and prominent emission-lines); and (3)  an unprecedented 
area of 15 deg$^2$ for a $z\simgt 1$ survey.  These give the CSI Survey a volume
comparable to the SDSS, with a selection method that efficiently traces stellar mass over 2/3 
the age of the universe ($0.4<z<1.5$) --- spanning the critical redshift range where cosmic
star formation precipitously drops, and groups and clusters become prominent.  
As the redshift survey with the largest unbiased volume at $z=1$,
CSI will allow us to
comprehensively address the interplay between environment, galaxy mass buildup, and star formation
at these redshifts.

The innovation of moving to low resolution for cosmologically interesting problems
of galaxy evolution was instigated by the PRIMUS collaboration
\citep[e.g.][]{coil2011,cool2013}. They designed a low dispersion prism to perform an optically selected redshift survey of several
legacy fields and have used their sample of 140,000 galaxies to constrain the evolution of galaxy stellar mass functions 
\citep{moustakas2013}, and to study AGN activity \citep{mendez2013} and it's relation
to galaxy evolution \citep{aird2012,aird2013}. 

In this, the sample definition and basic analysis paper of the CSI Survey, we describe the design of the project, with particular focus 
on the flux limits of the selection, the data processing and the spectral energy distribution (SED) fitting 
with its attendant determination of redshifts and redshift errors. We apply our methodologies to the first 
batch of data in 5.3 degs$^2$ of the SWIRE XMM-LSS field, and provide an overview of some basic 
properties of galaxies over the past 9 Gyr, with an emphasis on the gains made by selecting galaxies by 
stellar mass instead of by their rest-frame UV light.  


\section{The Carnegie-Spitzer-IMACS Survey}
\label{sec:survey}

\subsection{The limitations of optically-selected surveys}

Any effective probe of galaxy assembly must sample a wide range of masses in order to trace 
evolutionary connections between large and small systems. The build-up of massive red sequence 
galaxies may be driven by mergers with sub-$M^*$ systems, so it is essential to trace evolution to 
masses below $M^*$ to mitigate against the differential growth of the high- and low-mass 
populations. This trade-off between depth and area noted above has led to a dichotomy in 
redshift surveys. Narrow and deep programs, such as the Gemini Deep Deep Survey \citep[GDDS;][]{abraham2004} or
the Galaxy Mass Assembly Ultra-deep Spectroscopic Survey \citep[GMASS;][]{halliday2008}
could not cover enough volume to robustly sample the evolution of the high mass population at 
$z < 1$, while other surveys that have traded depth for area do not reach below $M^*$ with high 
fidelity. 

\begin{figure}[htb]
\centerline{
\includegraphics[width=2.5in]{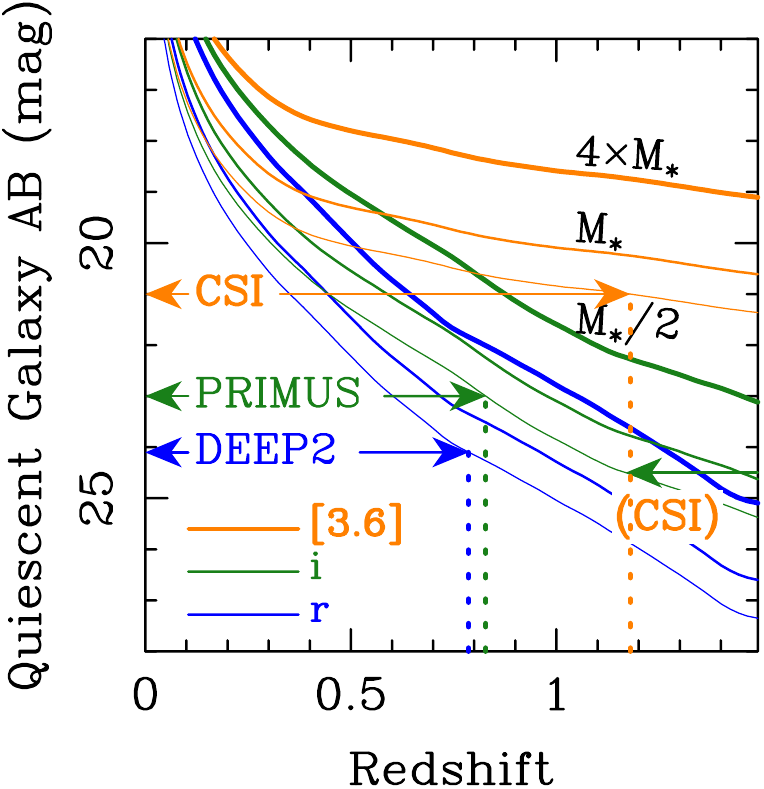}}
\caption{(left) Apparent magnitudes in $r$, $i$, and $3.6\mu$m as a
function of redshift, for passively evolving stellar populations ($3 \le z_f\le 6$) with
stellar masses $5M^*$ and $M^*/2$ \citep[$\log M^* \sim 10.85$ at $z \lesssim 1$;][]{drory2009}.
The magnitude limits for DEEP2 \citep{willmer2006} and PRIMUS \citep{coil2011} are
drawn. Such optical flux limits only cover the most massive passively evolving systems or those
galaxies with young unattenuated stellar populations, biasing galaxy samples at early times. The
shallow dependence of the $3.6\mu$m magnitude on redshift yields a selection with significantly less
bias against old systems. The effective optical limit of CSI, $i=24.5$ mag, is also shown.
The dotted vertical lines show the redshifts below which the DEEP2, PRIMUS, and CSI samples
are not biased against old galaxies with stellar masses $M=M*/2$.
\label{fig:masslimits}}
\end{figure}

An old stellar population at $z=1$ with a stellar mass of $10^{11}M_\odot$
corresponds to roughly $i=23$ mag and $r=24$ mag in the optical, as shown in Figure
\ref{fig:masslimits}.  The extreme optical faintness of such 
galaxies results in many of them being missed in optically-selected surveys despite
their relatively high masses. For example, the
selection limits for DEEP2 \citep{willmer2006} and PRIMUS \citep{coil2011}
are shown in this figure by the blue and green dashed lines, respectively. At $z>1$
these programs {\it required\/} galaxies to be extremely massive or have their optical light dominated
by young stellar populations in order to be fall within the survey selection.
At best it can be difficult to trace the evolution of the most massive galaxies with
such surveys. At worst, if not properly accounted for, the survey selection introduces biases and
systematic effects. Extremely deep optical limits can be taken as one valid approach to ameliorating
this problem, with the side effect of an overwhelming number of low-mass star-forming dwarfs
dominating one's source catalog.

\begin{figure*}[htb]
\centerline{
\includegraphics[width=4.5in]{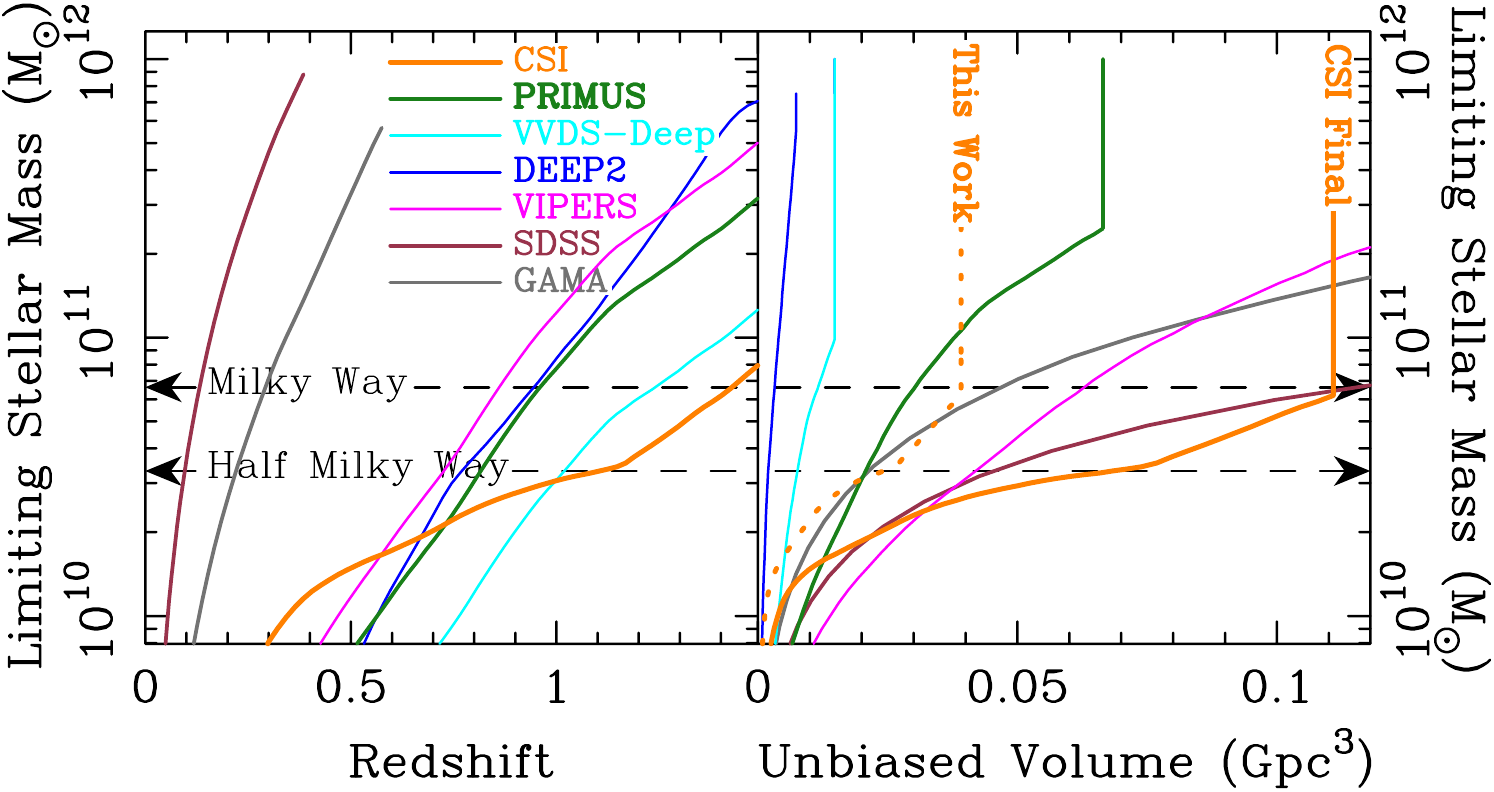}}
\caption{
(left) Limiting stellar mass of faint galaxy surveys by redshift. By using the \emph{IRAC}
$3.6\mu$m band, the CSI survey selection (the solid orange line) traces stellar mass more uniformly over most of the redshift range of the survey than samples selected in the optical.
Our resulting sample is currently reaching $i=24.5$ mag (dashed green line), probing down to stellar
masses equivalent to the present day Milky Way
\citep[$6.6\times 10^{10}M_\odot$;][]{mcmillan2011} out to $z=1.4$, almost order-of-magnitude lower than
DEEP2, and half the present day Milky Way at $z=0.9$, a factor of two deeper than DEEP2 or PRIMUS.
(right) Volumes probed with complete, unbiased samples for several redshift surveys as functions of limiting stellar
mass. When the 15 deg$^2$ is completed, CSI's volume (defined by the $3.6\mu$m- and $i$-band limits) will be more than an order of magnitude
larger than DEEP2. The volume traced by
the first 5.3 degs$^2$ (this paper) is shown by the dotted line.
\label{fig:masslimits_lim}}
\end{figure*}

\subsection{The potential of an \emph{IRAC}-selected survey}

The integrated light of all but the youngest stellar populations are dominated by light from 
the stellar giant branch, cool stars whose light output peaks in the near-IR.  It has long been
recognized \citep[e.g.,][]{wright1994} that this results in a $1.6\mu$m ``bump'' -- a peak in 
bolometric luminosity -- for galaxies with a wide range of star formation histories.
Indeed, as shown clearly in (for example) Figure 1 of \cite{sorba2010}, this feature is nearly unchanging in 
\cite{bc2003} models of stellar populations with mean ages 100 Myr $< \tau < $10 Gyr.  
Put in other terms, the near-IR mass-to-light ratio, M/L$_{1.6\mu m}$ changes slowly for all but the 
youngest stellar populations.  

The CSI Survey exploits the wide field and sensitivity of the \emph{Spitzer Space Telescope} 
with the \emph{Infrared Array Camera} (\emph{IRAC})  to take advantage of this property of the integrated near-IR light 
from stellar populations in $z\sim 1$ galaxies.  Combined with the insensitivity to internal and Galactic
extinction, selection at $3.6\mu$m closely mirrors selection by stellar mass.  Figure
\ref{fig:masslimits} directly compares the evolution of $3.6\mu$m magnitude with the optical $r$ and $i$
bands: the CSI selection wavelength has a dependence on redshift that is much shallower than surveys
selected in the optical. What slope remains for the CSI selection function is an unavoidable
\emph{k-correction}: over the redshift range $0.7 < z < 1.5$, the center of the $3.6\mu$m
\emph{IRAC} band
corresponds to restframe wavelengths of $2.1\mu m > \lambda_c > 1.4\mu m$, straddling the
$1.6\mu$m ``bump'' of the spectral energy distribution (SED) as it redshifts through the bandpass.  

Thus massive galaxies exhibit a much flatter trend of observed magnitudes with redshift
in the IR than in the optical, and this weaker dependence on galaxy mass afforded by $3.6\mu$m-selection of the sample is a 
key feature of the CSI Survey. Our goal has been to make a spectrophotometric survey to characterize
galaxy populations and environments in an unbiased way down to a stellar mass
of $M^*$ out to $z=1.5$, and down to a stellar mass of $M^*/2$ out to $z=1.2$.
As can be read from Figure \ref{fig:masslimits},
a mass limit of $M^*$ corresponds to $r=26$ mag or $i=24.5$ mag at $z=1.4$, and
$M^*/2$ corresponds to $r=26$ mag or $i=24.5$ mag at $z=1.2$.
Our current spectroscopic reduction and analysis, described below, is reaching
an effective optical limit of $i=24.5$ mag; brighter than this limit the CSI
spectroscopic sample is correctably complete (though not necessarily uniformly complete;
see \S \ref{subsec:success} for more details).

Figure \ref{fig:masslimits_lim} (left) plots the limiting mass as a function of redshift for CSI (solid orange)
and other redshift surveys. The depth of CSI in stellar mass is
substantially less sensitive to redshift compared to the others due to the \emph{IRAC} $3.6\mu$m selection,
varying by a factor $\sim 3$ over $z=0.5-1.5$, compared to $1-2$\,dex over the same redshift
range for optically-selected surveys. Consequently,
CSI samples a more uniform range of stellar masses over the full volume of the survey,
and to a lookback time of 9 Gyr.

This critical point is illustrated in Figure \ref{fig:masslimits_lim} (right),
plotting the depth in stellar mass against the volume probed with complete, unbiased samples.
DEEP2, shown in violet (the line colors are the same as in the left), is limited in both area
and mass depth. PRIMUS's 9 degs$^2$ is limited in depth, and thus only probes an
unbiased volume comparable to our first 5 degs$^2$. When CSI reaches its goal of 15 degs$^2$, the
survey will cover an unbiased volume equal to the SDSS with similar depth in stellar mass.
The large areas available from legacy Spitzer \emph{IRAC} surveys --- both wide {\it and} deep, and the
freedom from foreground and internal extinction at this wavelength, 
allows the construction of uniformly deep, homogeneous photometric samples for spectroscopic follow-up.

CSI reaches factors of 2-6 deeper than DEEP2 in mass, over an area
ultimately 8 times wider, and redshifts sufficiently accurate to characterize
environments by directly identifying groups and clusters. With such data we
aim to make the first group catalog at $z=1$ that is comparable in limiting mass and volume to SDSS,
and thus enable detailed environmental characterizations of galaxies at a time when the
universe was less than half its current age.

\begin{figure*}[htb]
\centerline{
\includegraphics[width=6.0in]{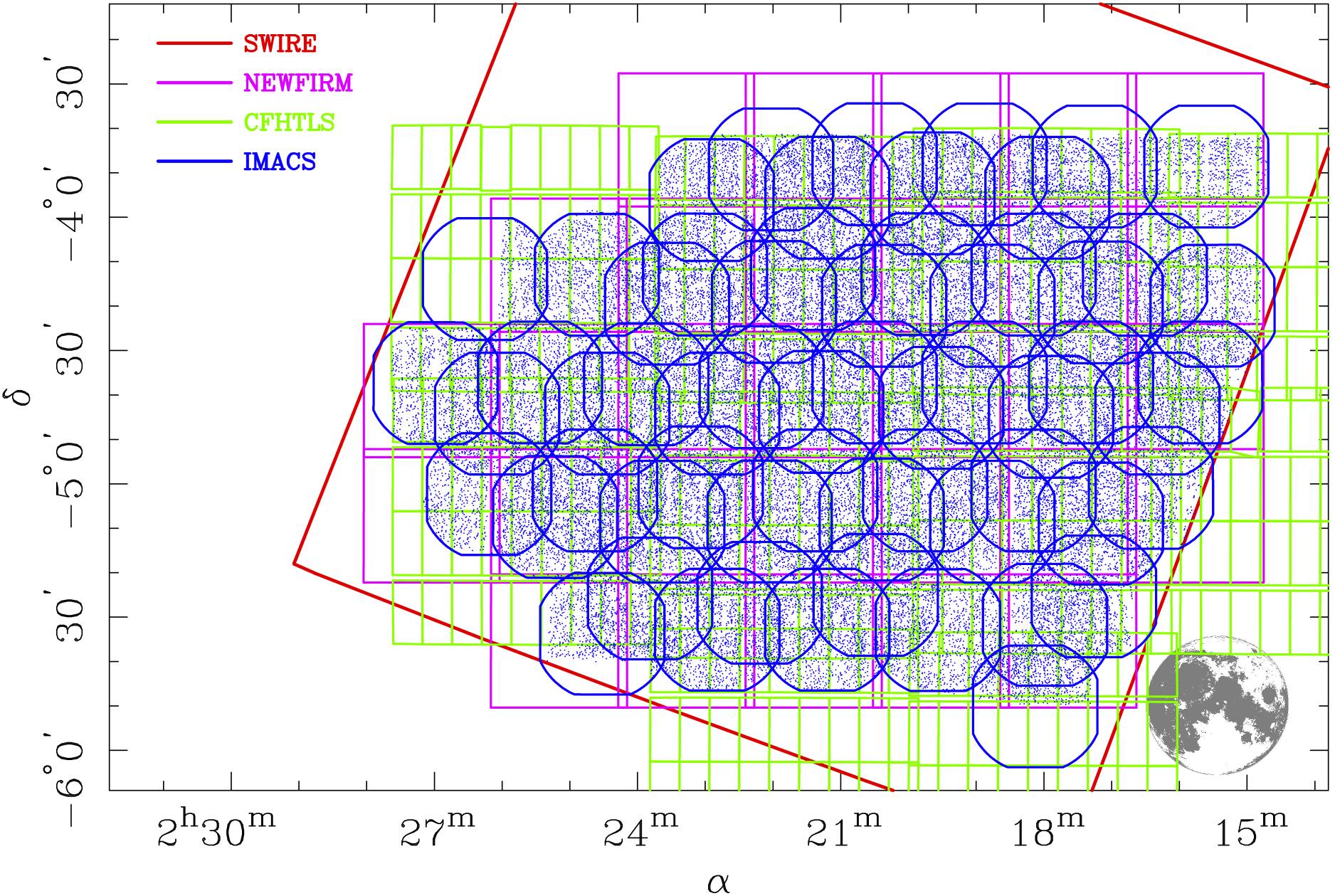}
}
\caption{A schematic of observations in the first 5 degs$^2$ of the SWIRE-XMM field
studied by the Carnegie-Spitzer-IMACS survey. The SWIRE \emph{IRAC} imaging field is outlined in
red. The CFHTLS-W1 optical data reanalyzed by us is shown in green. The first set of
NEWFIRM $J$ and $K_s$ observations are marked in violet. The positions of the IMACS
slitmasks are outlined in blue, with the blue points marking the positions of the 46,600
galaxies with CSI redshifts. The field of view of IMACS is comparable to the size of the
full moon, illustrated in the lower right.
\label{fig:xmmfield}}
\end{figure*}

In this first paper, we describe the details of our data reduction and redshift
fitting, redshifts for the first 45,284 galaxies in 5.3 degs$^2$ of the
SWIRE-XMM field (Figure \ref{fig:xmmfield}), and an initial characterization of the general galaxy populations
in our stellar mass-limited sample.

\section{Data}
\label{sec:data}

A project of this type and scope requires attention to detail and care in the processing and combination of a broad range of
data from different sources. In this section we describe the imaging data and reductions that underpin the 
broadband flux measurements for the sample, both for the ultimate goal of fitting SEDs, but also for
defining the selection criteria and incompleteness functions. Unless otherwise specified, all object 
detection was performed using SExtractor \citep{bertin1996}.

\subsection{Photometry}
\label{subsec:photometry}

The SWIRE Legacy Survey \citep{lonsdale2003} observed several large fields with \emph{IRAC} to a depth
suitable for our survey. Three of these fields are accessible from southern telescopes, providing up
to 23.8 deg$^2$ of 
coverage at $3.6\mu$m. Basic properties of the three fields are given in Table \ref{tab:fields}.
Of these fields, $15.3$ deg$^{2}$ have supplemental optical imaging
that is publicly available.
Of the 9.1 degs$^2$ of \emph{IRAC} imaging in the XMM-LSS field,
6.9 degs$^2$ is covered by the CFHT Legacy Survey
W1 $ugriz$ imaging. In this section we discuss our analysis of the SWIRE XMM-LSS \emph{IRAC}
data, our reprocessing of the CFHTLSW1 $ugriz$ imaging and subsequent  photometry of the $3.6\mu$m
catalog, as well  as our observations at $J$ and $K_s$ of the field using NEWFIRM \citep{autry2003}
on the Mayall 4m telescope at Kitt Peak National Observatory.

\subsubsection{Spitzer-\emph{IRAC} Imaging}
\label{subsubsec:IRAC}

The SWIRE Legacy Survey was a program undertaken to trace galaxies by stellar mass back to
$z=2$ \citep{lonsdale2003}. We obtained the
\emph{IRAC} images of the XMM-LSS field from the archive at IPAC. The reductions
and processing of these data were described by \cite{surace2005}. In order to 
minimize contamination of the object catalog by artifacts around stars, we configured SExtractor so as to ignore elliptical 
regions around bright stars. A `mexican hat' convolution kernel was used for object detection, resulting in 585,159 
objects in the $3.6\mu$m catalog of the XMM-LSS field. Fluxes were measured in a manner described by
\cite{surace2005}. Down to our selection limit of $3.6_{AB}=21$ mag the
$3.6\mu$m catalog contained 266,621 objects.

\begin{deluxetable}{l c c l}
\tablecaption{Basic Properties of the Southern SWIRE Fields Targeted by CSI
\label{tab:fields}}
\tablehead{
\colhead{Field} &
\colhead{SWIRE Area} &
\colhead{SWIRE+Optical} &
\colhead{Filters} \\
&
\colhead{(degs$^2$)} &
\colhead{(degs$^2$)} \\
}
\startdata
XMM-LSS & 9.1 & 6.9 & ugriz\\
ELAIS S1 & 6.9 & 3.6 & BVRIz \\
CDFS & 7.8 & 4.8 & ugriz
\enddata
\end{deluxetable}

\subsubsection{The Optical Imaging}
\label{subsubsec:optical}

The CFHT Legacy Survey \citep{cuillandre2006} targeted multiple fields using the one-degree MegaCam imager around 
the sky with a broad range of scientific goals. The ``Wide'' W1 
field provided almost 7 deg$^{2}$ of overlapping $ugriz$ coverage in the SWIRE XMM-LSS field. Unfortunately, the processed 
images and catalogs available at Terapix were not entirely suitable for our purposes, owing to uncertain astrometry and 
inadequate defringing of the $i$ and $z$ data. To remedy this issue, we obtained the complete set of calibrated frames from the CFHT 
archive and processed the data using the following additional steps. We begin by constructing new fringe frames in $i$ and 
$z$ by medianing scaled exposures obtained within a single night. These were then rescaled and subtracted from the 
individual exposures. Some bright objects did bias these medians when the number of exposures was small (i.e. $N \lappr 10$), 
leaving faint traces in the resulting fringe frames.  In general, however, the greater sky uniformity after subtraction of these new 
fringe frames improved the depth of our resulting catalogs by $\sim 0.2$ to $0.5$ mag compared to the Terapix catalogs. We 
constructed sky frames as well for the $u, g$, and $r$ bands using a similar methodology, though not restricting the 
construction to data collected within single nights.

New astrometric solutions were derived for each exposure using the \emph{IRAC} catalog as a set of deep astrometric standards. 
Cubic solutions for each chip were derived with a typical RMS scatter of 0\Sec 15. For those regions of the MegaCam images that
did not overlap the SWIRE data, we supplemented the catalog with the 2MASS point source catalog \citep{skrutskie2006}, in order 
to prevent the solutions from diverging at the edges of the \emph{IRAC} field.

Using these new astrometric solutions, and the photometric solutions from the headers of the individual frames, we constructed 
cosmic-ray-cleaned image stacks with 0\Sec 185 pixels, 30 arcmin on a side and evenly spaced. These smaller images were more easily 
managed than the larger image formats provided by Terapix, which allowed us to distribute the image analysis and photometry 
tasks over multiple processors. The zeropoints were checked by comparing the photometry of moderately bright objects with the 
data in the catalogs from Terapix and we found systematic offsets less than $\pm 0.03$ mag in every case.
Object catalogs were
generated using SExtractor, and these were matched to the \emph{IRAC} catalog with a tolerance of
1\Sec 5 arcsec, with the goal
of using the optical image characteristics to aid in star/galaxy separation, and to determine which \emph{IRAC}-selected objects 
have centroids with optical centers falling far from the defined slit positions.

\subsubsection{The Near-IR Imaging}
\label{subsubsec:nearir}

NEWFIRM \citep{autry2003} is a wide-field ($27^\prime \times 27^\prime$) near-IR camera deployed by NOAO at the Mayall 4m at Kitt Peak from 2007 to 
early 2010. During the fall semesters we imaged the XMM-LSS field in $J$ and $K_s$. Typical exposure times were
70 min/pixel in $J$ and 32 min/pixel in $K_s$. The seeing ranged between $0.8 - 1.4$ arcsec (FWHM).

The data were processed using a fully automated, custom pipeline, written as a prototype for a wide-field imager being deployed at
Magellan. The basic steps in the reduction were:
(1) subtraction of the dark current;
(2) correction for nonlinearity;
(3) division by a flat-field;
(4) masking of known bad pixels;
(5) construction of first-pass sky frames;
(6) derivation of image shifts, in arcsec, on the sky;
(7) stacking of the first-pass sky-subtracted frames;
(8) generation of a deep object catalog;
(9) construction of object and persistence masks for second-pass sky estimation;
(10) temporary interpolation over objects, persistence, and bad pixels for each frame;
(11) constructing bivariate wavelet transforms of these masked frames;
(12) fitting the time variation of the wavelet transforms of the sky using
shorter frequencies for larger spatial scales;
(13) reconstruction of model sky frames by inverting the temporally smoothed wavelet transforms;
(14) subtraction of the model sky frames;
(15) re-derivation of image shifts, in arcsec, on the sky;
(16) final stacking of the sky-subtracted frames, including the generation of sigma and exposure maps;
(17) identification of objects in the 2MASS point source catalog, restricted by 2MASS PSC quality flags;
(18) application of rotation, translation, and scale to the camera distortion based on the 2MASS objects; and
(19) calculation of image zero-points using the 2MASS objects.

Using the zeropoints of each image, we constructed image mosaics $30'$ on a side centered at the locations of
our $ugriz$ mosaics, but with 0\Sec 4 pixels.

\subsection{Aperture Photometry}

Magnitudes in $ugrizJK_s$ were derived using aperture photometry within a range of circular apertures for the
entire SWIRE catalog.  Because the seeing varied with wavelength and pointing, we convolved the $z$-band 
image stacks (which had the best PSFs on average) with Gaussians to simulate the effects of poorer seeing in the $ugriJK_s$ data. The offsets in 
magnitude from the degraded $z$ images were applied as PSF-corrections to the $ugriJK_s$ aperture magnitudes. 
While the details of the PSF and potential blending are critical for small apertures and for modeling the profiles of 
the objects, our choice of fitting the SEDs to much larger-aperture ($D=4"$) magnitudes allows us to utilize a simple and economical 
approach to the PSF corrections --- this choice was especially important given the enormous size of the object catalog. The choice of such a large aperture also reduces
the systematic errors in matching the aperture magnitudes to the \emph{IRAC} $3.6\mu$m fluxes.

\begin{figure}
\centerline{
\includegraphics[width=3.3in]{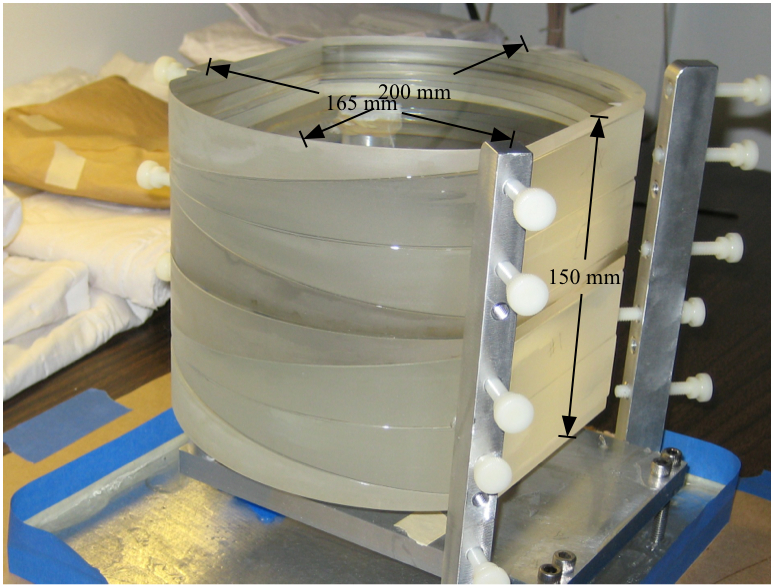}
}
\caption{Construction of the new ``Uniform Dispersion Prism,'' built for CSI.
Dr. S. Shectman designed the UDP to have a resolution of $R\sim 25$
from 7500\AA\ to 1$\mu$m. The eight layers are made from thin prisms of S-FPL51 and N-KZFS2, to make a stack
of glass 150 mm thick. This prism has a resolution with a mild dependence on wavelength (see Figure \ref{fig:resolution})
compared to the first prism deployed in IMACS \citep[see][]{coil2011}.
\label{fig:udpphoto}}
\end{figure}

\begin{figure*}[t]
\centerline{
\includegraphics[width=7.0in]{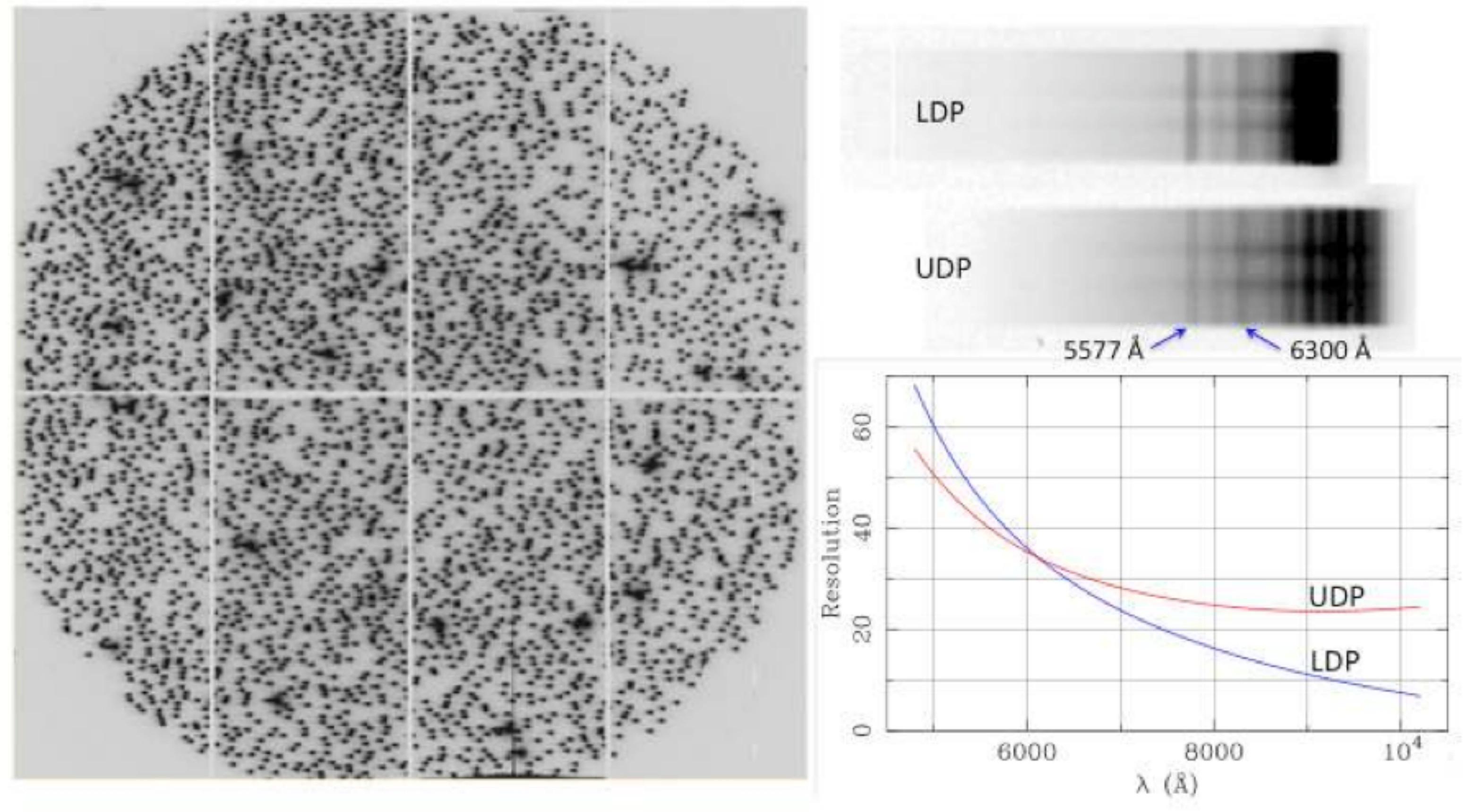}
}
\caption{\citep[Figure reproduced from][]{dressler2011}
(left) Multi-slit mask designed for use with the Low Dispersion Prism.
(top right) Example LDP spectrum --- nod \& shuffle \citep{glazebrook2001} produces
doubled object and sky spectra. Night sky lines [O I]5577\AA\ and [O I]6300\AA\ are
marked. Below the LDP spectrum is one obtained with the UDP, with its more uniform
dispersion (bottom right).
The resulting difference in resolution is shown in the bottom right. The resolution
provided by the new Uniform Dispersion Prism, designed for CSI, provides an improved
ability to trace the 4000\AA\ break and [OII]3727\AA\ to $z=1.4$ when coupled with IMACS's
red sensitive E2v detectors.
\label{fig:resolution}}
\end{figure*}

\subsection{Spectroscopy}
\label{subsec:spectroscopy}

While the concept of using
a low-dispersion prism for galaxy evolution has origins in PRIMUS, the infrastructure and 
background work for CSI's observing strategy and data reduction was largely developed during the first author's participation in the study of the
$z=0.83$ cluster RXJ0152--13 and the intervening field \citep{patel2009a,patel2009b,patel2011,patel2012}.
Between 2008 and 2009 the SWIRE XMM field was targeted in 31 multislit mask exposures with the \emph{IMACS} 
f/2 camera configuration and the Low Dispersion Prism (LDP) designed by S. Burles for the PRIMUS redshift survey
\citep{coil2011}. In 2010 we observed 29 SWIRE XMM masks using an innovative eight-layer disperser called the
Uniform Dispersion Prism (UDP, see Figure \ref{fig:udpphoto}).
Although both of these dispersers produce spectra with median resolutions of $\lambda/\Delta\lambda \sim 30$, 
the dispersion curve of the UDP is much flatter, providing significantly higher resolution
than the LDP out to $1\mu$m, at the
expense of marginally lower resolution in the blue compared to the LDP (Figure \ref{fig:resolution}).

Previous work by the PI on a similar program had obtained adequate $S/N$ ratios with exposure times of 3 hours 
down to $z'=23.3$ mag using less sensitive SITe detectors \citep{patel2009a,patel2009b,patel2011,patel2012}. Significantly more 
sensitive e2v detectors were installed in IMACS, thanks to support from the NSF's TSIP program (see Dressler 
\etal\ 2011); these CCDs boosted the throughput of the instrument by factors of 2-3 in the far red. We observed 
with these detectors using the Nod and Shuffle mode of IMACS
\citep[see][]{glazebrook2001}, and accumulated exposure times of 2h per galaxy for more than 90\% of
the sample, using individual integrations with durations of 30 min (15 min per position).
Typical slit lengths were 5 arcsec, with slit widths of 1 arcsec. The positions of the slit masks are shown in
Figure \ref{fig:xmmfield}. No blocking filter was necessary, as all of the light dispersed by
prisms stays within a single order.

Spectrophotometric standards were obtained during most observing runs, with a consistency of 5\% from run-to-run. Helium 
calibration lamps illuminating the deployable flat-field screen were taken approximately every hour (i.e., every other science 
exposure). In this section we describe the basic steps involved in reducing the multislit prism data using the custom pipeline 
that had been written for the earlier cluster programs
\citep[e.g.][]{patel2009a,patel2009b,patel2011}.

\subsubsection{Wavelength Calibration}

Two critical aspects of the reduction of LDP and UDP data involve mapping wavelengths on the detector,
and transposing between sky coordinates of objects and CCD coordinates.
The small number of He lamp lines are insufficient for determining accurate
wavelength solutions in individual slitlets, so we derive
global wavelength mappings over an entire CCD using all of the slitlets simultaneously. Thus, lines
in a single spectrum that may be corrupted by bad columns or pixels will not affect the solution for
that slitlet, as the $\sim 3000$ helium lines over a CCD frame constrain the fit.

The following are the basic, automated steps we have implemented in our pipeline: 
\begin{itemize}
\item Derive mapping of sky coordinates to CCD coordinates using an image of the slitmask;
\item Identify lines and fit 2D wavelength solutions using isolated helium lamp exposures on a chip-by-chip basis;
\item Refit new centroids for blended helium lines using nonlinear simultaneous Gaussian fitting on a slit-by-slit basis;
\item Fit for improved mapping of 2D wavelength solutions;
\item Shift the wavelength maps using 2-3 night sky emission lines (e.g., [O I] 5577\AA, Na 5890\AA, [O I] 6300\AA);
\end{itemize}

We start with a
distortion map for the camera, defined from previous exposures of the field around the globular cluster Palomar 5, to compute 
the approximate mapping of sky coordinates to CCD coordinates for the objects in the multislit mask. SExtractor is run on a 
direct image of the slit mask and a simple pattern analysis matches up the predicted positions of the slits and the measured
positions of the slits to create an adjusted mapping of sky coordinates to CCD coordinates.

With a theoretical
dispersion curve for the prism, we generate a trivariate mapping of sky coordinates to CCD coordinates that is wavelength 
dependent. SExtractor is run on a helium comparison lamp image and the resulting catalog of the 3,000 to 4,000 helium lines 
is pattern-matched to the predicted positions of 8 unblended helium lines (out of the 11 lines in our line list). New wavelength 
mappings are solved for as the order of the fit is gradually increased. The nearly final solutions generated at this stage are 
first order rescalings of the theoretical dispersion, with coefficients that are cubic polynomials of the sky coordinates. 
The typical RMS at this stage is 0.3-0.4 pixels. With these solutions, we can now generate predicted positions for the 
complete helium line list and fit Gaussian profiles to the full line list for each slit. Within each slit, the Gaussians are fit 
simultaneously, providing accurate positions for both the unblended and blended helium lines. Using these new positions for 
the helium lines in CCD coordinates, we derive through iteration, the trivariate mapping between CCD coordinates and
both wavelength and sky positions, with a typical final RMS scatter of 0.25 pixels. For LDP data this scatter corresponds 
to $\sim 50\,$\AA\ at  $9000\,$\AA, or $\delta z \sim 0.005$. This large uncertainty in the wavelength calibration in the far red for the LDP is 
due to the fact that its dispersion reaches $\sim 200\,$\AA/pixel at $9000\,$\AA; the scatter of $\sim50\,$\AA\ introduces a 
25\% uncertainty in $\Delta\lambda$ at such wavelengths when computing the flux calibration. Because the resolution with the 
UDP is $3\times$ higher at these red wavelengths, this additional source of uncertainty is negligible for those data.

The final step involves cross-correlating simulated night sky emission lines ([OI]5577\,\AA, Na I 5890\,\AA, and [OI]6300\,\AA)  
placed at their predicted positions with the spectra in the science frames, on a slit-by-slit basis. The median offset in both $x$ 
and $y$ is then applied to the wavelength mappings found from the nearest helium lamp exposure.

\subsubsection{Extracting Spectra}

Due to the unique nature of these prism data, the science frames must also be
processed in a nonstandard fashion using custom written routines. We take particular care in the 
extraction of spectra because of the small number of pixels covered by each object, the faint optical limits we expect to probe, 
and the proximity of the objects to slit ends where residual sky counts can bias simple extraction algorithms.

As can be seen in Figure \ref{fig:resolution} (top right), each object is exposed twice per readout.
In our implementation of nod \& shuffle, the object positions are shifted by 1.6 arcsec between the two
spectra, A and B. Due to the fairly rapid timescale of the nodding (typically 60s), the sky
backgrounds in the A and B exposures are functionally identical, and the subtraction of the A and B
two-dimensional spectra from each other should leave only the object remaining. However, while nod \& shuffle
does provide excellent sky subtraction, the galaxy spectra themselves still must be divided by a flat-field.
Flat-fielding data from nod \& shuffle is also complicated by the fact that the A and B spectra were both
exposed on identical detector pixels. Thus the pixels in the flat-field defined by the position of the A spectrum should
be used to flatten {\it both\/} the A and B spectra. To summarize the first two steps of our reductions,
first subtract 2D spectrum A from 2D spectrum B, and B from A. Next, divide the A-B spectrum and the B-A
spectrum by those pixels in the (normalized) flat-field that were covered by spectrum A.

A first attempt to extract 1D spectra for every object in an exposure involves a simple form of optimal
extraction \citep{horne1986}, in which each object is assumed to have a Gaussian spatial profile.
The centroid of the Gaussian is defined as a low-order polynomial of wavelength, and this polynomial defines the
initial trace of the object. The second moment of the Gaussian is also computed as a low-order polynomial function of
wavelength. Then a rough 1D spectrum for a given object can be obtained using this Gaussian approximation for
the spatial profile in a manner similar to \cite{horne1986} but we opt to solve for the 1D spectrum in a
least-squares sense, representing it as a b-spline \citep{dierckx1993}, and allowing us to downweight bad pixels,
cosmic-rays, or other discrepant data.
With this approximate spectrum, we solve for new parameters of the spatial profile, including
up to 4 Hermite moments (10 for spectrophotometric flux standards), with each of these moments defined as
low-order polynomial functions of wavelength. The wavelength binning of the b-spline is defined by
the wavelength solution for each slitlet.

When these spatial profiles and object moments have been derived for all
eight CCDs of a given exposure, we analyze 
the differences between the predicted and measured object positions. These deviations are well described by a linear 
function of coordinates in the slitmask, largely reflecting the small misalignments of the mask
that occurred when the observations were taken. We use this 
new linear function to fix the object positions, re-compute spatial profiles, and re-extract 1D spectra for each
object in each exposure.

After deriving the traces and Gauss-Hermite moments of all the objects in all the exposures of a given slit mask, we
perform a final optimal extraction using all of the exposures of a mask simultaneously. This procedure allows us to
better flag (and ignore) cosmic-rays and other bad pixels, and to use the improved 1D spectrum of each object to
further refine the spatial profiles and object positions. In this final extraction that combines all of the
exposures, we incorporate the sensitivity function from our flux-calibration and produce spectra
in physical units. Pixels, now all converted to flux units, are weighted according to the inverse of the
expected noise and bad columns and pixels flagged as discrepant are given zero weight.
Iteration allows us to flag and reject most cosmic rays, and to also re-weight the pixels using the ``least-power'' M-estimator
(see below) in an effort to discount the noise estimates initially derived from the data frames. Blindly using noise
that has been estimated from the original frames can lead to subtle, small biases in the resulting spectra; the
iterative re-weighting helps mitigate the issue.
The process performs an optimal extraction that yields
a single 1D spectrum that matches all the data in a least-squares sense. While the exposure times are fairly
uniform, the sky noise can vary greatly from exposure to exposure, especially as one approaches twilight or
nearer to the moon. Given the small differences between the wavelength solutions of the exposures, we sample the
one-dimensional b-spline representation of the spectrum at a fixed set of wavelengths, defined by the dispersion
of the prism at the center of the field. This fixed grid greatly simplifies the construction of templates for
fitting the SEDs.

Spectra for objects repeatedly observed in overlapping slitmasks were not combined at this stage, but kept separate --- for
this paper --- so that we could better assess data quality and empirically estimate our redshift precision.
Approximately 18\% of the galaxies were repeated; we discuss these below in the context of the redshifts and
their uncertainties. Approximately 20\% of the objects observed with the LDP were also 
reobserved with the UDP.

\begin{figure*}[t]
\centerline{
\includegraphics[width=6.5in]{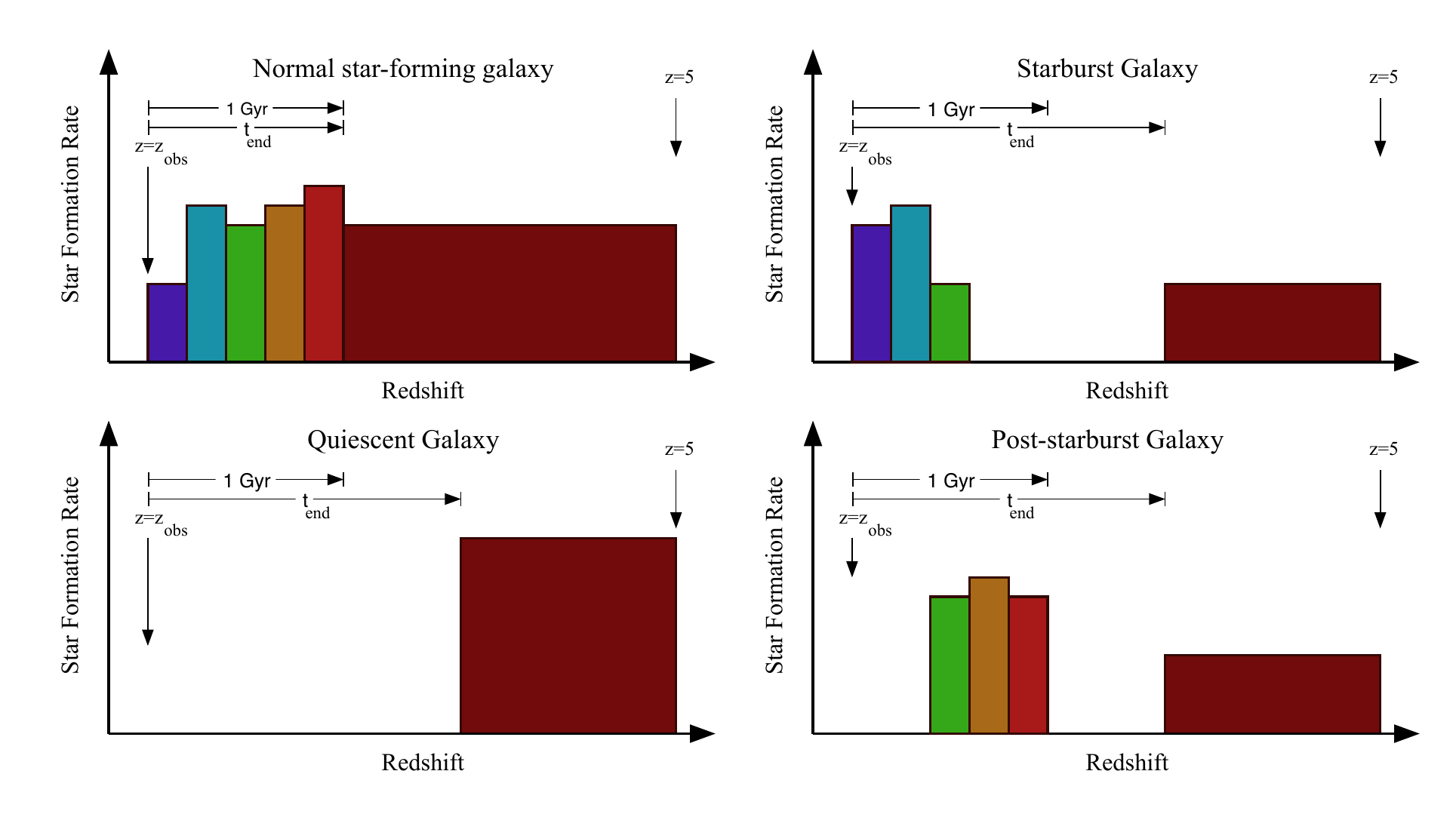}
}
\caption{Cartoon star formation histories for four 
representative galaxy types. Because optical passbands provide poor leverage on 
star formation histories earlier than 1 Gyr prior to the epoch of observations, we have reduced the complexity of galaxy star formation 
histories to nonnegative combinations of discrete components such as those illustrated in these cartoons. For our purposes, each galaxy is 
modeled with six age-related components, as described in the text, with the oldest component starting at $z=5$ and continuing down to 
some time $t_{end}$ prior to the epoch of observation. Five younger, discrete components of duration 200 Myr allow for the broad possible 
range of complex histories in a galaxy's recent past. Each of these six components is quadrupled, with four different levels of extinction, 
$A_V\in \{0,0.5,1,2\}$ mag, leading to a total of 24 stellar components with non-negative contributions to the stellar mass of a galaxy. With 
redshift, metallicity, and $t_{end}$ as our gridded parameters, there are 24 stellar coefficients at each location in the grid. By abstracting 
star formation histories in this way, we have retained the essential information content of on-going star formation, intermediate-age 
populations, and an old, underlying stellar population. 
\label{fig:cartoon}}
\end{figure*}

\subsubsection{Spectrophotometric Flux Calibration}

Accurate flux calibration is a critical component of the survey because spectral
slopes provide additional information to constrain galaxy redshifts. For CSI we choose 
bright hydrogen white dwarf standard stars with deep,
broad Balmer lines that can be detected with the LDP/UDP. 
When taking a standard star spectrum, we place it in the center of an 
alignment star box on one of our slitmasks to ensure we are capturing essentially all of the star's
flux. 
Deep Balmer lines are thus required since they allow us to refine  
the wavelength calibrations for the stars when they are not accurately centered in the boxes.

The processing of standard star exposures is similar to that described above for the 
science exposures, with the 
following modifications. 
In order to derive the sensitivity function, we must first compensate for the fact that the seeing in the standard star exposure leads to 
different resolution than one would obtain with a narrow slit. We make an assumption that the seeing was isotropic and that the Gauss-Hermite 
parameterization of the spatial profile is a valid descriptor of the seeing profile in the dispersion direction. Using this profile as a kernel, we 
deconvolve the extracted spectrum using an implementation of {\sc CLEAN\/} \citep{hogbom1974}, convolving the result to a resolution 
equivalent to that defined by the science data (FWHM=4.5 pixels).
The wavelength calibration of the alignment star box is calculated with the trivariate wavelength
solution of a helium lamp exposure obtained immediately after the standard star, with an additional
correction from the measured positions of the star's Balmer lines. 

Due to time and weather constraints, spectrophotometric flux standards were not obtained during every night, but
the excellent run-to-run repeatability of the relative flux calibration from 8500\,\AA\ to 4500\,\AA\ for the
LDP (and from 9000\,\AA\ to 4500\,\AA\ for the UDP), of approximately $\pm 5\%$
allows us to use flux standards obtained on different nights or during different runs.
We correct for slit losses, aperture corrections, and other such systematics as part of our SED fitting procedures, discussed in the
next section.

\section{SED Fitting}
\label{sec:fitting}

With broadband photometry and spectra in hand, a suitable library of template spectra can now be employed to estimate 
spectrophotometric redshifts. In this section we describe the basis functions in our templates, implement a generalized maximum 
likelihood method and estimate confidence limits for the redshifts of each galaxy
along with several other parameters and properties.

Here we discuss the construction of the templates, using several continuum components, each derived from the \cite{maraston2005} models. 
The \cite{kroupa2001} initial mass function (IMF) was used, resulting in a median offset from
``diet'' Salpeter IMF (Bell \& de Jong 2001) of 0.04 dex for SSP ages up to about 9 Gyr.

\subsection{Ingredients}
\label{subsec:ingredients}

Ideally, a set of spectral templates should not only span the redshifts probed by a survey, but also encompass the potential 
range of optical and near-IR properties of galaxies over those epochs. Satisfying this latter constraint cannot be done {\it a priori\/} 
as it is one of the chief goals of the project. However, models of evolving stellar populations have been used for modeling 
low-dispersion prism data for the purpose of recovering redshifts to $z\sim 1$ \citep[see][]{patel2010}. Our
method for CSI echoes that approach, and we describe it here.

Our templates are constructed as the superposition of several continuum components and multiple emission lines. The stellar population 
bases were derived from the \cite{maraston2005} models, using the
\cite{kroupa2001} initial mass function (IMF)\footnote{This is equivalent 
to applying an offset of -0.04 dex to the stellar $M/L$ ratios of ``diet'' Salpeter IMFs (Bell \& de Jong 2001) for single stellar population 
(SPP) ages up to about 9 Gyr.}.  In building our templates we exploit the well known fact that for times $\tau \gappr 1$ Gyr after the 
cessation of star formation, optical data no longer provide significant leverage on a galaxy's prior star formation history \citep{tinsley1972}. 
We adopt a simple set of stellar population basis functions: these components are sensitive to different timescales of a galaxy's star 
formation history; in combination, they reproduce the broad range of SED properties seen in normal galaxies.

The first base component is a constant star formation model with a starting epoch of $z_f=5$. The time at which star formation ceases for 
this component is a free parameter in our analysis, a grid of values ranging from $10^9$ yr to $10^{10}$ yr prior to the redshift of observation 
(capped at $z=5$), whose values are spaced logarithmically with an interval of 0.10 dex. The metallicity of this base population is also 
gridded with values ranging from [Z/H]$=-1.2$ to [Z/H]$=0.6$, with an interval of 0.2 dex. Redshift is the final gridded parameter, ranging 
from $z=0.005$ to $z=2.0$ at intervals of $\Delta z=0.005$.  Superimposed on the base population are up to five piecewise constant SFR populations
that formed in 200\,Myr intervals centered on lookback times of $t=1,3,5,7,9\times 10^8$ yr,
with the same metallicity as that of the base. Four cartoon star 
formation histories built out of such components are schematized in Figure \ref{fig:cartoon}.  The SEDs of each of these stellar components 
are redshifted, convolved to the wavelength-dependent resolution of the instrument, and sampled at the wavelengths appropriate for the 
galaxy spectra.  Lastly, extinctions of $A_V\in \{0.0,0.5,1.0,2.0\}$ mag are applied to each component in an effort to reproduce
the effects of complex dust distributions that are not well modeled by simple screen models.

In addition to the continuum components, we include several emission line components: (1) a single unresolved Gaussian emission line 
for [OII]3727\,\AA, (2) a blend of three Gaussians at [OIII]5007\,\AA, [OIII]4959\,\AA, and H$\beta4861$\,\AA, with ratios of $1:1/3:1/10$, to 
broadly mimic the typical line ratios seen in galaxies at these masses \citep[e.g.][]{kauffmann2004}, (3) a Gaussian for H$\alpha6563$\,\AA, 
and (4) an unresolved emission line for MgI2799\,\AA.  Because of the increasing uncertainties in the flux calibration of LDP data beyond 8000\,\AA, 
no emission lines were allowed beyond that point in those fits. Given the higher quality of the UDP data in the far red, we extended this limit 
to 9000\,\AA\ in fitting UDP data, with the exception of [OII]3727\,\AA, which we accepted to 9400\AA.  These restrictions 
do not greatly affect the redshifts that are measured, but they do have some impact on the fit to the red continua of low redshift galaxies. 
Note that the widths of the Gaussian profiles are fixed to that defined by the spectral resolution.

Thus there are in total 24 stellar continuum components and 4 emission line components, each redshifted, convolved to the 
wavelength-dependent resolution of the instrument, and sampled at the wavelengths appropriate for the galaxy spectra. The broadband 
magnitudes of each component are computed using the filter transmission and detector QE curves of the respective imaging instruments.

\subsection{Fixing the Flux Calibration}
\label{subsec:fixing}

Before solving for the 28 coefficients at every location in the three-dimensional grid of redshift, metallicity, and termination time,
we perform a step that solves for wavelength-dependent slit losses and aperture corrections, so that the IMACS
spectrum and broadband photometry can be fit together as a single SED.
This rescaling of each IMACS spectrum is derived using the broadband $griz$ photometry from the 4 arcsec diameter
apertures. Performing this task on a galaxy-by-galaxy basis can be done accurately and robustly when the $S/N$
ratio of the photometry exceeds that of the spectroscopy, such as was done with deep SuprimeCAM data in \cite{patel2010}.
But in CSI, the CFHTLS-W1 data
has a 5-$\sigma$ depth of $i\sim 24.5$ mag, and fitting for a separate wavelength dependence for each galaxy is
not robust. Therefore we use the bright, red galaxies observed in a given mask to define
a simple polynomial dependence for such slit losses and aperture corrections. The procedure is as follows:
(1) construct a reduced set of templates, (2) fit these templates to the broadband fluxes of the bright, red galaxies,
(3) derive rough photometric redshifts, (4) use these best-fit photometric templates to construct template
prism spectra, and (5) ratio these prism templates with the observed prism spectra and fit
low order polynomials. The resulting function is used for the
wavelength dependence of the aperture corrections/slit losses for all objects within a given mask, allowing for
a simple scaling of this polynomial for every galaxy to match the prism
spectra with the broad-band photometry. At low redshift the adoption of a single polynomial to
fix the flux calibration for an entire mask
breaks down for large galaxies, contributing
to some incompleteness at bright magnitudes. No significant correlations of the residuals from this fitting procedure
are seen with respect to the positions of objects on the slit mask.

\subsection{Fitting for the Galaxy Components}
\label{subsec:fitting}

Standard least-squares and $\chi^2$ minimization techniques are susceptible to bad data points and outliers, for example, from a few residual 
cosmic rays or unidentified bad pixels, especially given the small number of usable pixels in each spectrum ($\sim 150$). We opted to perform 
an iteratively reweighted non-negative least squares fit for the template coefficients at each location in the grid, using the weight 
function of Huber's M-estimator \citep{huber1981,zhang1997}. Doing so is equivalent to minimizing Huber's M-estimator itself,
given as $L_{Huber}$, with the associated weight function $W_{Huber}$:
\begin{eqnarray}
L_{Huber}(x) =
&\begin{cases}
L_2(x) & \text{if $|x| \le k$}\\
k(|x|-k/2) & \text{if $|x| \ge k$}
\end{cases}\\
L_{2}(x) = & x^2/2\\
W_{Huber}(x) =
&\begin{cases}
1 & \text{if $|x| \le k$}\\
k/x & \text{if $|x| \ge k$}
\end{cases}
\end{eqnarray}
where $k=1.345$. Near the optimum location, points retain the behavior of $L_2=x^2/2$ (i.e. least squares) while outside points retain the 
behavior of $L_1=|x|$ (i.e. minimum absolute deviation), leading to an estimator that is robust against outliers but sensitive to the details of the 
distribution close to the optimum location. In three iterations the routine effectively minimized this M-estimator.  By minimizing $L_{Huber}$, instead of the $L_2$ 
M-estimator, the fitting produces a set of coefficients that are robust against outliers and 
contaminated data, so long as the fraction of bad pixels is $\lappr 25\%$.

Using the rescaled IMACS spectra and the $ugrizJK_s$ photometry, we solve for the template coefficients using iteratively 
reweighted non-negative least squares. The coefficients for each galaxy are stored, as well as several M-estimators for each grid location, for use 
in the final likelihood analysis (described below).

\begin{figure*}[t]
\centerline{
\includegraphics[width=6.5in]{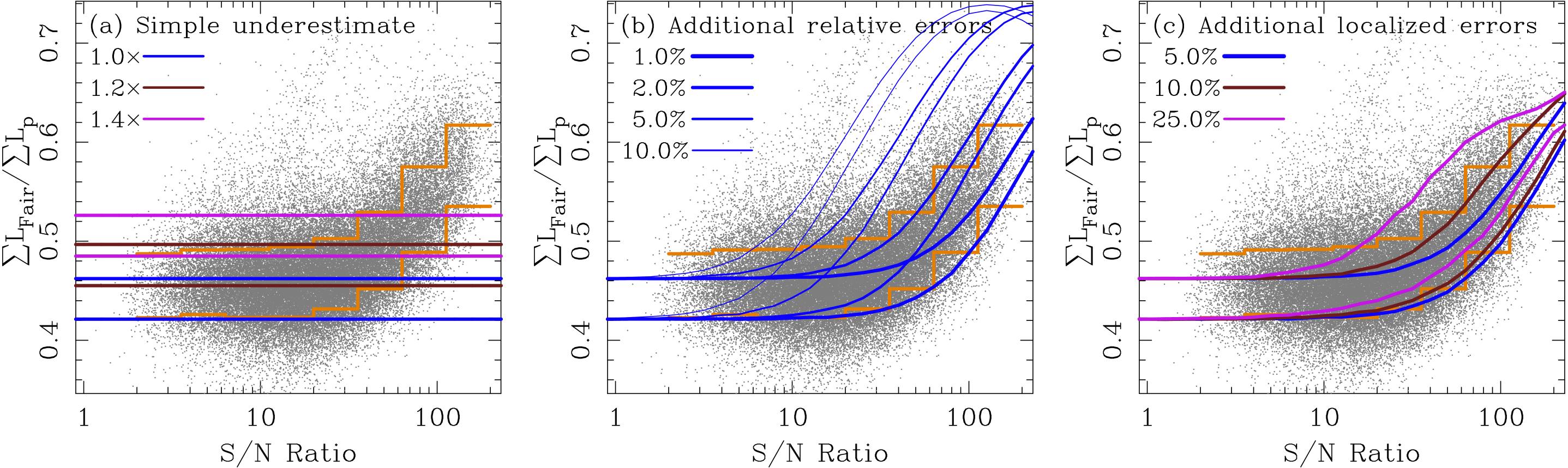}
}
\caption{Ratio of M-estimators $L_{Fair}$ and $L_p$ (see text) vs signal-to-noise ratio of the
IMACS spectra (gray points). Orange lines show the 16th and 84th percentiles of the CSI data in
bins of $S/N$. (a) Simple models where the true noise is $1\times$, $1.4\times$, and
$2.0\times$ that used when performing a fit. These models show that are low and moderate $S/N$
ratios the CSI noise estimates have not been significantly under-estimated, but at high $S/N$
ratios there are additional sources of flux errors. For each model family we show a pair of lines to
illustrate the 16th and 84th percentiles of the distributions.
(b) Simple models are shown with additional (Gaussian) noise added at levels of
1\%, 2\%, 5\%, and 10\% of an object's flux, proceeding from thick to thin. The curves mimic
relative sources of random error such as simple flat-fielding uncertainties.
The 16th and 84th percentiles are shown.
(c) Models with a 1\% ``flat-fielding error'' and added
(Gaussian) noise contaminating 5 pixels per spectrum at levels of 3\% (green), 12\% (blue), and
40\% (violet). Pairs of lines, again, show the 16th and 84th percentiles. Taken together, these models imply
that the CSI spectra are largely consistent with $\sim 1\%$ flat-fielding or other relative errors and
additional, localized sources of noise. In other words, residuals from the SED fitting show that the noise 
in the spectra is indeed non-Gaussian.
\label{fig:Mrat}}
\end{figure*}

\subsection{Likelihood Analysis}
\label{subsec:likelihood}

Each location in the three-dimensional grid of redshift, metallicity, and the termination of star-formation for the base `constant-star-formation' 
model has a dozen coefficients, as well as goodness of fit metrics. With these grids of population coefficients and M-estimators, confidence
limits on all of the gridded parameters and derived properties, such as total masses, stellar masses, restframe colors, were
calculated using a likelihood function
\begin{equation}
\begin{split}
d\mathcal{L} = &dz d\log t_{end} d[Z/H] P(z,\log t_{end},[Z/H])\times\\
&\exp\bigl[{-\sum L_{Huber}(x|z,\log t_{end},[Z/H])}\bigr]
\end{split}
\end{equation}
where $P$ is the defined set of priors at a given redshift, termination time, and metallicity,
and $x$ is the set of observations over which the M-estimator is summed.
In our likelihood analysis we found that using
Huber's M-estimator gave results that were most robust to bad pixels without biasing the contributions of emission lines, though others, for
example, ``least-power'' or ``Fair'' \citep[see, e.g.,][]{zhang1997}, were also very effective. We caution against using M-estimators that 
are insufficiently concave, as pixels ``contaminated'' by real (but narrow) features such as unresolved emission lines are attributed less weight 
than continuum pixels (by definition).
With these three dimensional likelihood functions, we have 24 SFH + 4 emission line parameters at each location 
in the grid.

For every spectrum, we marginalize over redshift, stellar mass, rest frame
colors, stellar population parameters as per the components in the SED fitting, and emission
line luminosities, estimating 68\%, 90\% and 95\% confidence intervals for
every parameter and property. Several criteria are employed to define the sample that has
successfully passed through the SED fitting, including sensible restframe colors and
physically meaningful stellar masses. We also restrict the sample to the set of galaxies with
formal 95\% uncertainties in rest frame $M_g$ of $\Delta_{95}M_g < 2.0$ mag (equivalent to
$\sigma_{M_g}< 0.5$ mag). The result is a relatively clean sample of 45,286 galaxies with a
total of 55,490 spectra, over a field of 5.3 degs$^2$.

We will use these confidence limits, along with data quality
measures derived from additional M-estimators, to identify a high quality sample
in the next sections.

\subsection{Using M-estimators to Probe Data Quality}
\label{subsec:quality}

Assessing the quality of any data involves more than the estimation of $S/N$ ratios and calculations of
reduced $\chi^2$ --- statistics that rest on simplistic or potentially unrealistic assumptions for
the underlying sources of noise. In this section we investigate the distributions of residuals from
the SED fitting using M-estimators and construct a new statistic to assess data quality.
Because they display varying degrees of sensitivity to outlying data
points, comparisons of M-estimators directly probe how well the errors in the flux
measurements have been estimated. Because one's measurement uncertainties underpin any likelihood analysis,
confidence in one's measurement errors translates directly to confidence in any confidence
intervals. And while no simple combination of
M-estimators will provide a complete description of the distribution and sources of bad data,
the heuristic approach shown below provides important initial insights that we expect to push further as we
complete the remainder of the survey.


We first assume that on average the residuals from the maximum likelihood analysis reflect the true
distribution of flux errors. The magnitude of these residuals, normalized by our estimated
measurement errors, probes both the extent to which uncertainties have been improperly estimated and
the extent to which bad pixels contaminate the data. Recall that most likelihood analysis is of the form
$\exp{(-L_2)}$, with some limited analysis of $\chi^2$. However, such work can only
provide insight so long as $[f_{model}(\lambda)-f_{observed}(\lambda)]/\sigma(\lambda)$ is
drawn from a Gaussian distribution with a standard deviation of unity.
M-estimators have been designed to mitigate against specific non-Gaussianities --- long tails and
sporadic spurious points --- and such ``best fit'' solutions are significantly less biased by bad
data. However, wholesale under- or over-estimation of measurement errors may remain an issue in
one's data, and different M-estimators will reflect this to varying degrees.
Thus, ratios of M-estimators have the potential to
become powerful tools for identifying galaxies in our sample that have been contaminated by
spuriously bad pixels, have simply had their flux errors under-estimated due to poorly handled
systematics, both intrinsic to the data or introduced in the reduction process, or have substantial
systematic errors due to template mismatch or poor calibration (e.g. flat-fielding).

For this analysis we introduce the ``Fair'' and ``Least-power'' M-estimators \citep{zhang1997}:
\begin{eqnarray}
L_{Fair}(x) = & c^2\bigl[ \frac{|x|}{c} - \log(1+\frac{|x|}{c})\bigr]\\
L_{p}(x) = & |x|^{1.2}/1.2
\end{eqnarray}
and use the ratio $\sum L_{Fair}/\sum L_{p}$ as a means to assess data quality.
A higher ratio indicates that the formal errors have been underestimated.
Since this ratio doesn't include $\sum L_{Huber}$, which was optimized in our SED fitting,
the ratio $\sum L_{Fair}/\sum L_{p}$ provides a relatively independent check on the data quality.
This statistic represents a helpful moment of the distribution of residuals from the SED fitting in a way
that captures the extent to which the noise was properly accounted for. After all, the noise estimates
underpin the shape of $\mathcal{L}(z)$.

For a Gaussian distribution of residuals, with perfect estimates for the measurement uncertainties,
the ratio $\sum L_{Fair}/\sum L_p$ converges to $\sim 0.44$. For a survey with many thousands of
spectra with only 50 resolution elements, 68\% of $\sum L_{Fair}/\sum L_p$ should fall between $0.42
\simlt \sum L_{Fair}/\sum L_p\simlt 0.46$. For non-Gaussian distributions the ratio increases.
Therefore, a finding of $\langle \sum L_{Fair}/\sum L_p\rangle\gg 0.46$ for a set of data indicates
that the noise estimates were underestimated --- though one does not know if the
elevation in noise was due to contamination by spurious data or to a simple underestimate of the Gaussian
noise. But by using simple Monte Carlo experiments we calculate how this ratio is correlated with
$S/N$ ratio under the assumption that a range of systematic measurement errors plague the data set.
Comparing the resulting trends with what is seen in the CSI data, we can better understand how to
divide the CSI sample according to a quantitative assessment of data quality.
For this analysis we consider only the residuals from the fits to the spectra,
ignoring the broadband photometric residuals.

Figure \ref{fig:Mrat}(a-c) shows $\sum L_{Fair}/\sum L_{p}$ plotted against $S/N$ for those objects
that passed the initial SED fitting quality checks. Orange lines show the 16th and 84th percentiles of the data in bins of $S/N$
ratio. Several simple noise models are overlaid. In Figure \ref{fig:Mrat}(a) the blues lines mark
the 16th and 84th percentiles one obtains with Gaussian noise that has been perfectly estimated.
The maroon and violet lines represent ratios obtained when
the noise is underestimated by factors of $1.2\times$ and $1.4\times$, respectively.
The spectroscopic residuals from our SED fitting are not consistent
with such simple underestimates of the errors in the IMACS spectra.

In Figure \ref{fig:Mrat}(b), we show models in which additional (Gaussian) noise is added at levels of 1\%, 2\%,
5\%, and 10\% of an object's flux, with progressively thinner lines. These curves
indicate that our data have
additional sources of noise unaccounted for in our analysis that are typically at levels of a few percent or
less. Such levels may result from errors in flat-fielding or from a general mismatch of templates, for example.
However, the additional sources of noise need not be uniformly distributed within each spectrum. In
Figure \ref{fig:Mrat}(c), we start with the 1\% ``flat-fielding'' noise and contaminate 1 resolution element,
with additional, localized noise at levels of 5\%, 10\%, and 25\% of an object's flux in those pixels.

While the permutations of such models may be infinite, the residuals in the CSI data are reasonably well
described by some combination of the estimated noise due to the sky, object, and detector electronics, {\it
plus\/} flat-fielding errors at levels of $\sim 1\%$, {\it plus\/} additional sources of bad pixels containing
noise at levels of several percent. These
trends should remind reader the combined sources of noise in the data are indeed non-Gaussian, justifying
our use of M-estimators in this maximum likelihood analysis.

\begin{figure*}[t]
\centerline{
\includegraphics[width=4.5in]{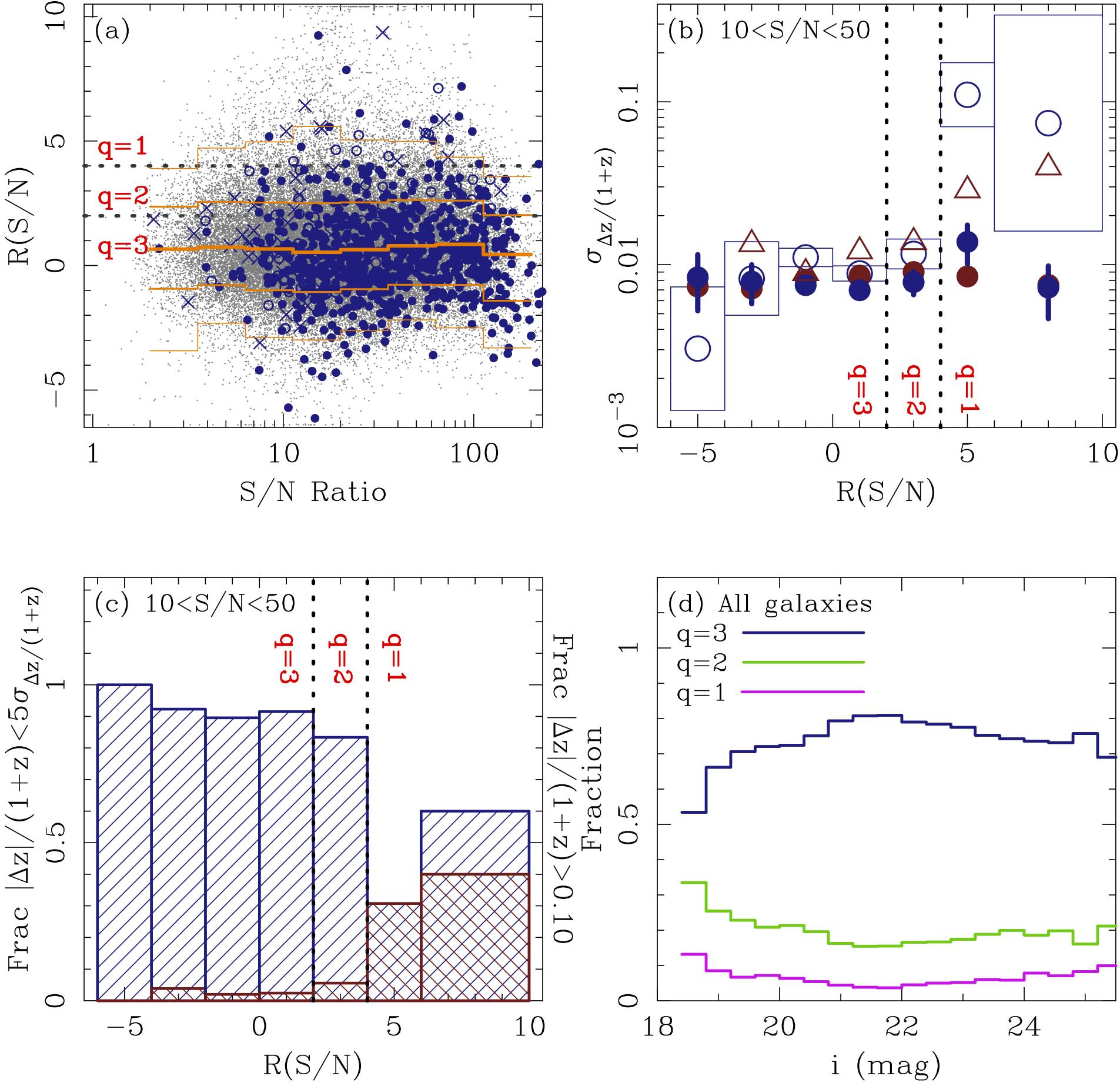}
}
\caption{
(a) The ratio $\sum L_{Fair}/\sum L_{p}$, renormalized by the
noise model shown by the blue lines in Figure \ref{fig:Mrat}(c), shown as a function of $S/N$
ratio. Large values of $R(S/N)$ reflect larger moments of the distributions in residuals from the SED fitting than
expected from the estimated errors.
(b) For galaxies with $10< S/N < 50$, we plot the robust standard deviation in the
fractional redshift errors of galaxies vs $R(S/N)$ using open circles, with the rectangles denoting the
1-$\sigma$ uncertainties. Filled blue circles mark the mean redshift uncertainty in each bin as
derived from the CSI likelihood functions. Red triangles show the estimated redshift errors derived from
those objects with repeat observations, while the filled red circles mark the mean redshift uncertainty
derived from the CSI likelihood functions for those particular galaxies.
Objects with $R(S/N)\simgt 4$, have significantly larger redshift errors than estimated from the likelihood
functions, and thus earn the lowest quality flag, $q=1$.
(c) The fraction of galaxies
with redshift errors less than $5\sigma$ [see the blue points in b] are shown with blue
hatched regions. The fraction of catastrophic failures ($>0.10$ in $|\Delta z|/(1+z)$) is shown in dark red.
Objects with $R(S/N)\simlt 2$, have catastrophic failure rates of $<5\%$. These define
the highest quality sample, $q=3$, in which $>90\%$ of these galaxies also have redshift errors $|\Delta z|/(1+z) < 5
\sigma_{\Delta z}/(1+z)$. Intermediate quality, $q=2$, is defined by $2<R(S/N)<4$.
(d) We plot the fractions of galaxies with each quality flag as a function of $i$-band
magnitude. The quality flags are not, per se, definitions with respect to absolute redshift
uncertainty but definitions regarding the quality of the redshift uncertainties themselves.
\label{fig:Mrat2}}
\end{figure*}

\begin{figure*}[t]
\centerline{
\includegraphics[width=6.0in]{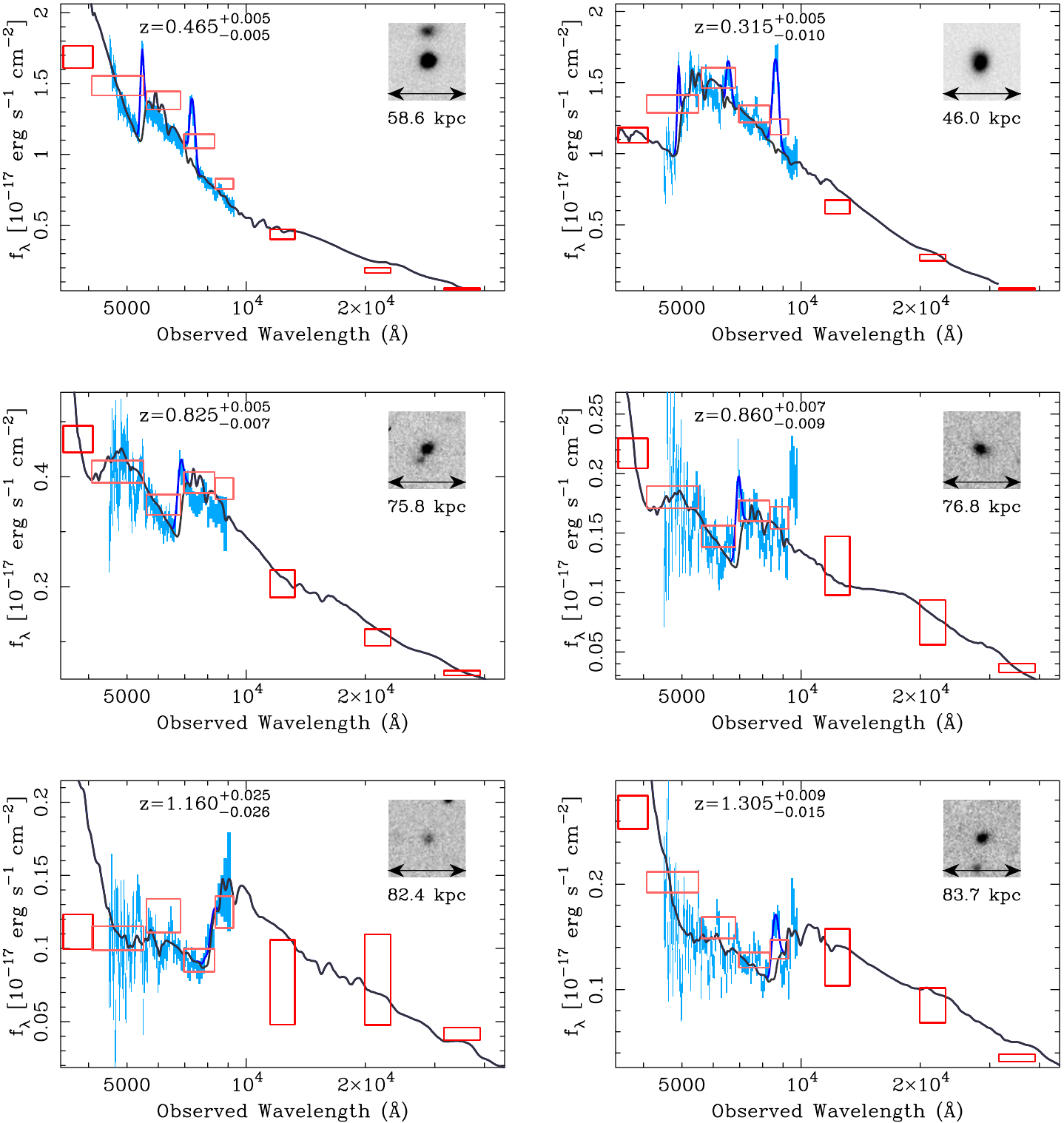}}
\caption{Example SEDs for blue galaxies, along with their CSI redshifts and 95\% confidence limits.
The IMACS spectra are shown in cyan. The red boxes mark the broadband flux measurements in
$ugrizJK_s[3.6]$, noting that the SED fitting was performed using $ugrizJK_s$. The best-fit stellar
population models are shown with thick black lines. When the data require emission line components
in the fit, these are shown using dark blue. Inset we show the CFHTLS $z$-band images of the
galaxies. In particular, we note that for a number of these galaxies the broadband
photometry alone would be insufficient to provide high quality redshift estimates.
Low dispersion spectroscopy was
key to isolating Balmer breaks and emission lines in blue galaxies, leading to
very little dependence of redshift uncertainties on spectral type in CSI.
\label{fig:examples1}}
\end{figure*}

\begin{figure*}[t]
\centerline{
\includegraphics[width=6.0in]{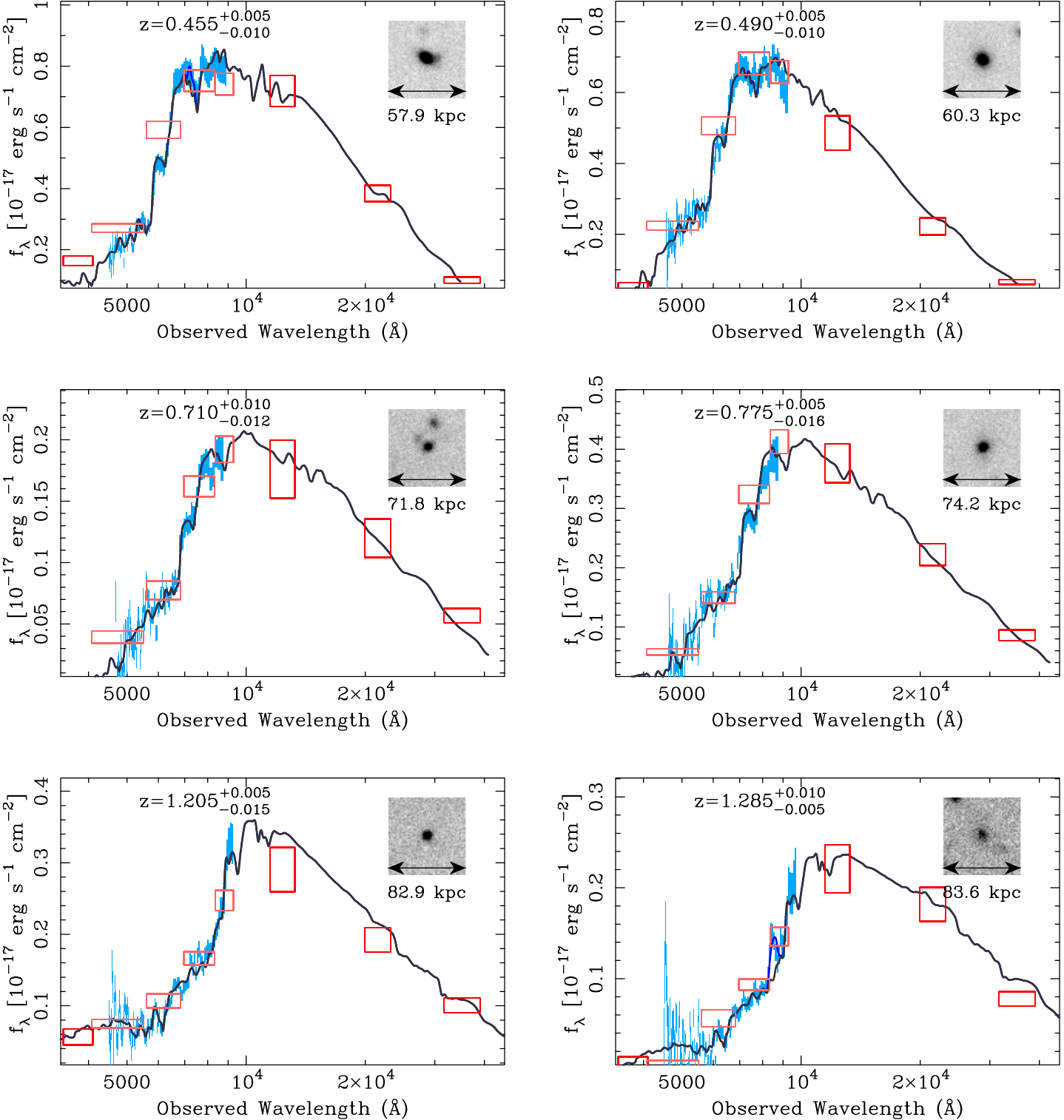}}
\caption{Same as in Fig. \ref{fig:examples1} but for red galaxies in CSI.
Note the structure in the SEDs traced by both the data and the models outside of the 4000\AA\ break.
\label{fig:examples2}}
\end{figure*}

\begin{figure*}[t]
\centerline{
\includegraphics[width=4.5in]{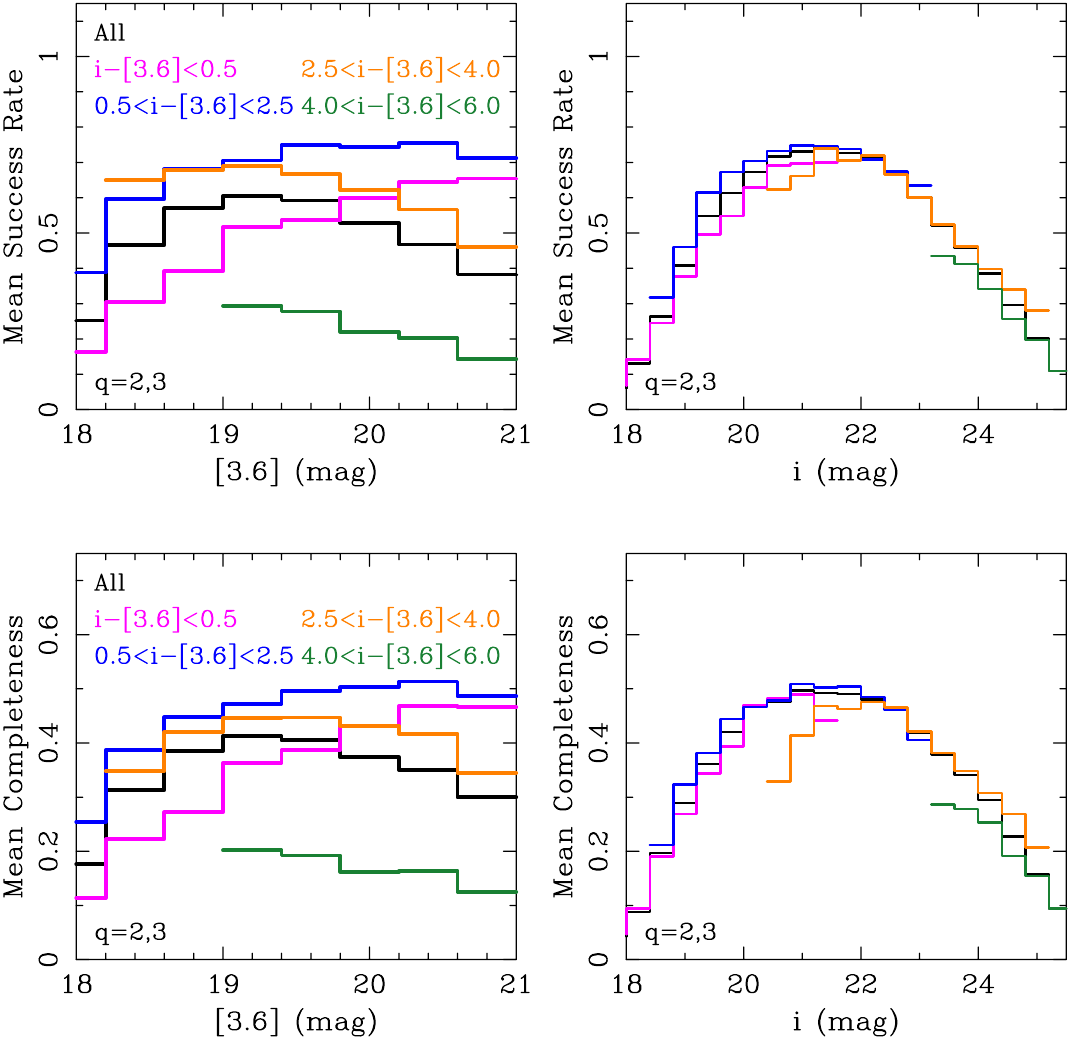}
}
\caption{(top) Average success rate
as functions of $3.6\mu$m and $i$ magnitude, where success rate refers to the fraction of slitlets cut that
led to high quality redshift measurements ($q\in \{2,3\}$).
(bottom) The average completeness
as functions of $3.6\mu$m and $i$ magnitude, where completeness is the fraction of galaxies
above the $3.6\mu$m flux limit within the 5.3 degs$^2$ area that have CSI
redshifts. At a given $3.6\mu$m flux, the completeness is a
fairly strong function
of color, such that it very nearly depends on $i$-band magnitude alone.
\label{fig:comp}}
\end{figure*}

\subsection{Defining Data Quality Flags}
\label{subsec:flags}

We now adopt the noise model shown in blue in Figure \ref{fig:Mrat}(c) as the baseline
for defining spectroscopic data quality, since it effectively follows the lower boundary of the locus
of points. We refer to the 50th percentile curve of this set of Monte Carlo
simulations as $M(S/N)$, and use the 68\% population interval, at fixed $S/N$, to
define the significance of a deviation from $M(S/N)$:
\begin{equation}
R(S/N) = \log\bigl[\frac{\sum L_{Fair}}{M(S/N)\sum L_{p}}\bigr]/\sigma_{\log M(S/N)}
\end{equation}
At a given $S/N$ ratio, galaxies that lie far above $M(S/N)$ either have problematic
data or have severely underestimated flux uncertainties. $R(S/N)$ thus serves as a diagnostic
of data quality based on the CSI data themselves, requiring no external redshift or spectral data.

Figure \ref{fig:Mrat2}(a) plots $R(S/N)$ vs $S/N$ ratio. The small gray points are the same as
in Figure \ref{fig:Mrat}. The orange lines trace the 2nd, 16th, 50th, 84th, and 98th
percentiles. The large, blue points are galaxies with previously known spectroscopic redshifts.
In this initial CSI sample, there are 710 galaxies with redshifts known from high resolution data.
Roughly half of these can be found in the VVDS \citep{lefevre2003}, another $\sim 90$ from
\cite{cooper2013}, and the remainder come from the UDS itself
\citep{simpson2010,akiyama2010,smail2008}\footnote{http://www.nottingham.ac.uk/astronomy/UDS/data/data.html}.
We have visually inspected and verified redshifts of the VIMOS spectra of galaxies common
to the VVDS and CSI, in a few cases revising VVDS redshifts and quality flags. For VVDS objects
flagged as `low quality' (3 \& 2), we were not able to verify all redshifts; we nevertheless
include them here. We will return in \S \ref{subsec:accuracy} to a more in-depth comparison of
redshift measurements once we have defined data quality flags.

In Figure \ref{fig:Mrat2}(b) we use the open blue circles to plot the standard deviation of the
fractional redshift errors in bins of $R(S/N)$, \cite[using $\sigma=1.48\times$ MAD, or
median absolute deviation;][]{beers1990}, using bootstrapping to estimate the 1-$\sigma$ uncertainties
marked by the rectangles. We restrict the sample to those
galaxies with spectra having $10<S/N<50$, in order to avoid being biased by any underlying
trends of $S/N$ ratio with $R(S/N)$. The blue filled circles show the mean fractional redshift uncertainty as
derived from our confidence limits on the redshifts. Galaxies with $R(S/N)\simgt 4$ have significantly larger
redshift errors than was estimated from our likelihood functions. For those galaxies with repeat measurements,
we estimate their mean redshift errors and plot those with red triangles. Red filled circles indicate the
expected mean redshift uncertainties for those galaxies and the repeat measurements reinforce our conclusion
that objects with $R(S/N)>4$ show redshift errors significantly larger than the uncertainties derived from the
likelihood analysis.

In Figure \ref{fig:Mrat2}(c) the blue hatched region shows the fraction of galaxies with
redshift errors less than 5-$\sigma$, where $\sigma$ is the mean estimated redshift error
from the confidence limits (i.e. the blue points in b). In red , we show the fraction of galaxies with
redshifts in error by more than $|\Delta z|> 0.10\times (1+z)$.
This ``catastrophic failure'' rate is $<5\%$ for galaxies with $R(S/N)\simlt 2$, and $<10\%$ for galaxies with
$R(S/N)\simlt 4$. With this in mind, we define a ``quality'' parameter, $q$, based on $R(S/N)$ cuts.

The highest quality CSI data are those galaxies with $R(S/N)<2$ ($q=3$).
For galaxies with $2< R(S/N)<4$ ($q=2$), the catastrophic failure rate remains low, at $5.5\%$.
We therefore consider
galaxies with $R(S/N)<4$ ($q=2-3$) as adequate for general scientific work. CSI data with $R(S/N)>4$
make up the low quality subset ($q=1$) of CSI. Note that while we restricted
Figures \ref{fig:Mrat2}(b) and \ref{fig:Mrat2}(c) to those galaxies with $10<S/N<50$,
the basic trends are consistent over both narrower and broader $S/N$ ranges.

Figure \ref{fig:Mrat2}(d) shows the relative fractions of galaxies with each quality flag as
functions of $i$-band magnitude. Given that the quality thresholds in $R(S/N)$ run nearly
parallel to the orange lines in Figure \ref{fig:Mrat2}(a), it is not surprising that the
fraction of the sample with a given quality flag does not depend strongly on magnitude
for galaxies fainter than $i=19$ mag.
At bright optical magnitudes, the spectroscopic residuals become dominated
by template mismatch (and even saturation in extreme cases). For the
rest of the paper, we refer to the ``highest quality'' sample as the set of 36,481 galaxies with
$q = 3$, and the ``high quality'' sample as the set of 43,347 galaxies with $q\in \{2,3\}$. 
For the remainder of the paper, we ignore the 1937
galaxies with $q=1$.

Example data and SED fits illustrating the ability of CSI to trace spectral features over a broad
range of redshifts and spectral types are shown in Figures \ref{fig:examples1} and
\ref{fig:examples2}.

\subsection{Success Rates and Completeness}
\label{subsec:success}

In a spectroscopic galaxy survey, the degree to which the population is sampled can be characterized
by two quantities: the success rate (fraction of \emph{targeted} galaxies for which usable spectra are obtained) and the
completeness (fraction of the original \emph{photometric} sample that is observed and usable spectra obtained).
The former quantity is naturally folded into the latter, which also involves fundamental observational 
limitations (finite telescope time, slit collisions, and imperfect detectors, among others).

In most high resolution spectroscopy, the successful measurement of a redshift  hinges on
the (visual) identification of specific features, often prominent emission and/or absorption lines.
To be considered a secure redshift in such surveys, multiple features typically must be seen.
A program's success rate, which can depend strongly on magnitude and/or color, is usually intertwined with the
underlying distribution of galaxy redshifts and emission line luminosities; ``successes'' occur when
these features fall within the spectroscopic
window with sufficient S/N ratios to be reliably identified.

On the other hand, incompleteness in photometric redshift surveys depends on 
fits of template SEDs to flux measurements made over a (hopefully) wide baseline in 
wavelength. Specific narrow features are not required to fall within a particular window; rather, redshifts are largely constrained by broad
breaks in
galaxy SEDs. At low $S/N$ ratios the SED fitting can degrade, 
sometimes leading to larger uncertainties in the derived parameters (redshift, age, etc), and 
occasionally resulting in unphysical solutions. Artifacts like poorly handled bad pixels, 
cosmic rays, improper fringe correction, and blending may similarly cause the SED fitting
to fail for some fraction of spectroscopic targets. 

The Carnegie-Spitzer-IMACS Survey, with its combination of low-resolution spectroscopy and broadband photometry,  
does not depend on the presence of critical spectral features within the spectroscopic bandpass. 
Instead, the factors which contribute to incompleteness in CSI more closely resemble those that affect 
broadband photometric surveys. The presence of,
e.g., strong [OII]3727\AA\ can certainly lead to highly secure redshifts in some cases. However, the power of this survey comes with the 
ability to trace intermediate-width
structure within the SEDs, such as the broad Ca H\&K absorption, the G band, or even the structure around the MgII
absorption at restframe 2800\AA\ in older stellar populations.

Success for CSI therefore depends on the quality of the photometry and the extracted 1D spectra, which jointly constitute our SEDs.
Having too low a S/N ratio in the IMACS spectra is usually an issue at magnitudes fainter than i=23.5, and at these
flux levels the photometry is especially crucial. The portion of a spectrum used
in the SED fitting is only about 140 pixels long, leaving them susceptible to 
bad CCD columns and artifacts at the ends of the short slitlets

\begin{figure*}
\centerline{
\includegraphics[width=5.0in]{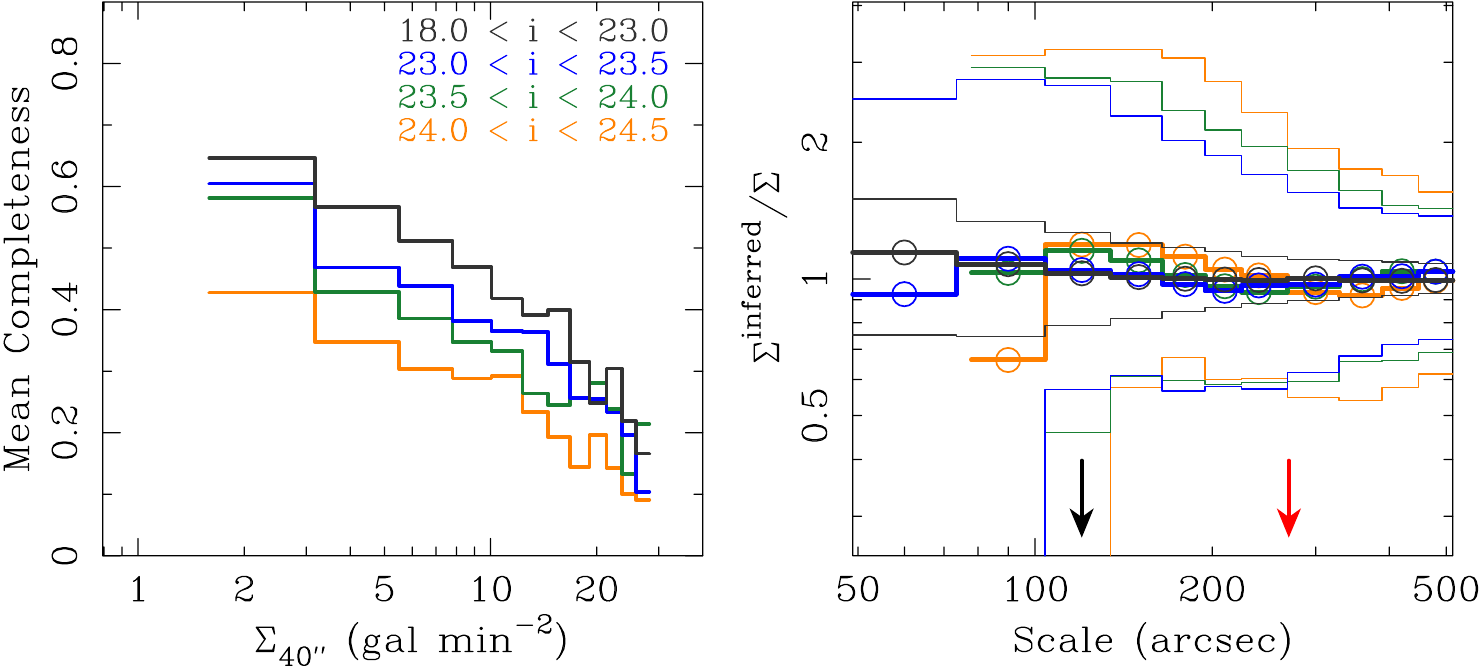}
}
\caption{(left) The mean completeness as a function of local source density computed on
scales of $40\times 40$ arcsec, approximating the portion of our incompleteness arising from slit
collisions. (right) The accuracy with which local galaxy counts are reproduced by our
high-quality sample alone, when summing the inverse of the completeness as weights on
individual galaxies. The thick lines indicate that, on average, there is no systematic bias
in the local source densities inferred from the CSI redshift catalog down to $i=24.5$ mag,
with a standard deviation of $0.1-0.3$ dex, depending on the scale over which galaxy number
counts are estimated.  
Using the scaling relations for massive halos from
\cite{carlberg1997} and \cite{carlberg2001}, one infers
that halos with masses $M_{vir} = 5\times 10^{14}M_\odot$ and $M_{vir} = 5\times
10^{13}M_\odot$ at $z=1$ have typical diameters of $\sim$ 4.5 arcmin (red arrow) and $\sim
2$ arcmin (black arrow), respectively. Based on these
calculations We infer that slit collisions and other sources of incompleteness introduce no
systematic errors in our accounting of the numbers of galaxies present in group-sized halos
and larger.
\label{fig:dens}}
\end{figure*}

In Figures \ref{fig:comp} (top) and (bottom) we plot the CSI success rates and
completeness as functions of $[3.6]$ and  $i$-band magnitudes, also breaking down the sample into coarse bins of galaxy color.
We model the completeness as a 2D function of both
$3.6\mu$m magnitude and $i-[3.6]$ color, and for illustrative purposes Figure \ref{fig:comp} shows the mean rates of success  and
completeness within broad color bins.
We are thus able to estimate
the incompleteness in the final sample given a galaxy's magnitude, color, and local source density.
Our peak success rate is 70\% between $20<i<22.5$ mag, and at those magnitudes the
completeness is about 50\%.
At $i=24$ mag the current pipeline produces a mean success rate of 40\%, resulting in a
mean completeness for CSI of 30\%. Down to $i\sim 24.5$ mag the incompleteness is well-characterized and can be corrected; below this, the
completeness corrections become more problematic (as does the CFHTLS photometry).  

To first order, the success rate is simply a function of $i$-band magnitude. Notably,
the distribution of estimated redshifts of the lower quality
data ($q=1$) is not statistically different from the high quality sample except at 
faint magnitudes ($i>23.5$ mag).  In any case, it is critical that our completeness corrections correctly model what was
lost as functions of magnitude, color, and most importantly, local source density.

This last factor profoundly influences the completeness of spectroscopic surveys like CSI through slit collisions. 
To model this, we divided our sample into bins based on local source density.  The footprint of a 
CSI spectrum is $\sim 200$ pixels wide, or 40 arcsec, by 10 arcsec (including both the A and B spectra).
For every target and observed galaxy, we measure the local source density in $40\times 40$ arcsec ($4/9$ arcmin$^2$) boxes. We then compute
the mean completeness in bins of local source density, shown in Figure \ref{fig:dens}(left).
This declining function illustrates the worsening effects of slit collisions at high projected source densities. We
also have compared the bivariate ($[3.6]$, $i-[3.6]$) completeness functions in coarse bins of source density, but find no
statistically significant variation.
This consistency allows us to treat separately
the decreasing completeness as a function of source density, and the joint effects of
$i$-band magnitude and $i-[3.6]$ color.

\begin{figure*}[t]
\centerline{
\includegraphics[width=5.0in]{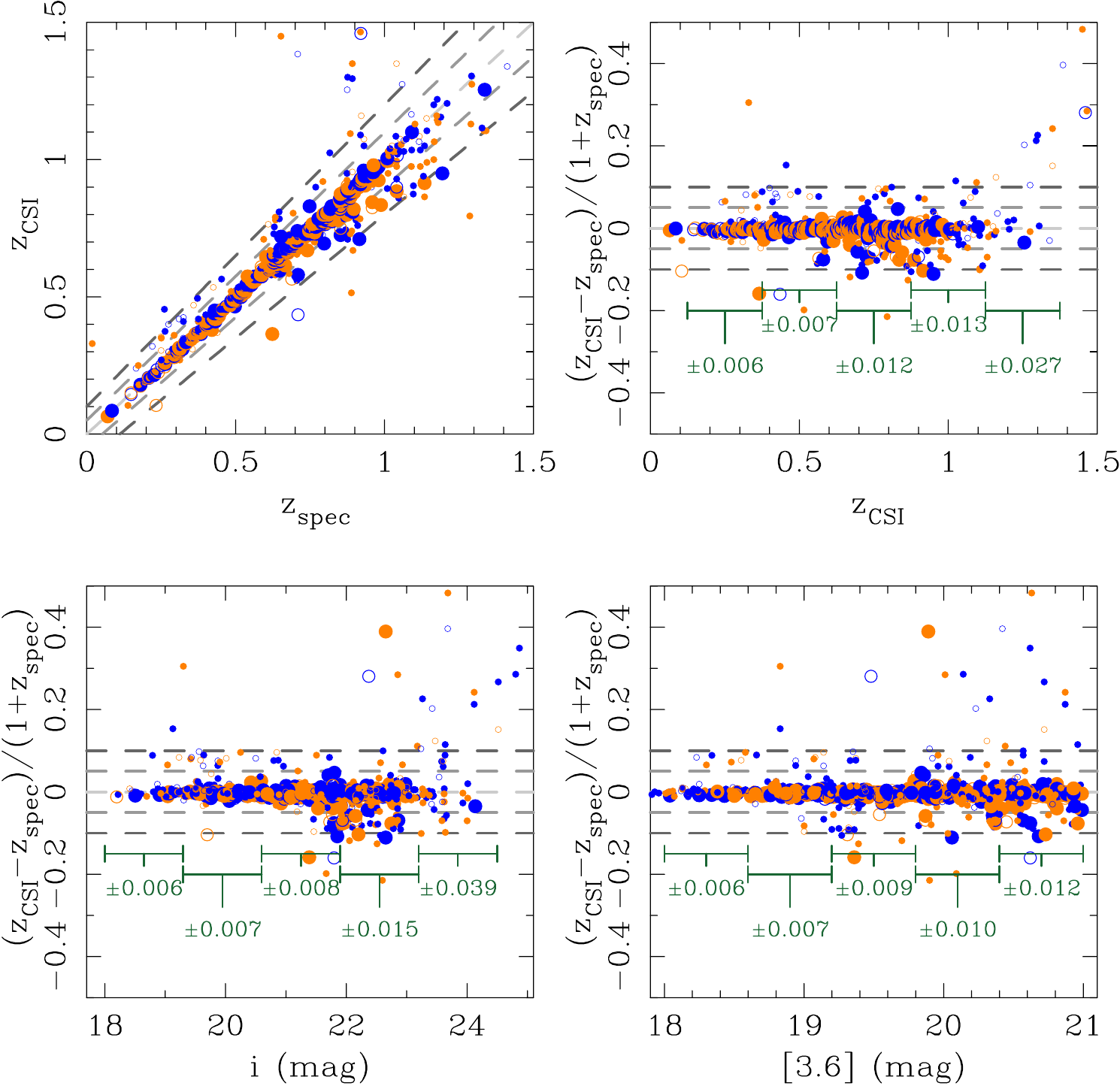}
}
\caption{(top left) Galaxies with high quality CSI redshifts ($q\in\{2,3\}$) plotted against redshifts in the VVDS
and in the UDS from \cite{cooper2013} with quality flags 3,4,5.
Larger symbols indicate higher quality redshift flags from the high resolution datasets.
The dashed lines mark the loci of $|\delta z/(1+z)| = 0,0.05,0.10$ from lightest to darkest, respectively.
In orange and blue we show
those galaxies observed with the LDP and UDP respectively. (top right) Fraction redshift
error vs. VVDS redshift. Numbers shown with bins along the x-axis show the $1\sigma$
estimate of the scatter, computed using $\sigma=1.48\times$ MAD \cite[median absolute
deviation;][]{beers1990}. Overall agreement is excellent, with a median systematic bias of
$\langle \delta z/(1+z)\rangle=-0.004$ and low scatter that is largely correlated with redshift and $i$-band
magnitude, and less so with $3.6\mu$m magnitude. For the 85 galaxies with CSI redshifts $0.7<z<1.0$ and high quality
VVDS redshifts (quality flags 4 \& 5), 95\% have redshift differences $\delta z/(1+z)<0.05$ and a $1 \sigma$ scatter of
$0.010$ in $\delta z/(1+z)$.
The numbers of CSI measurements with redshift errors
$|\delta z/(1+z)| > 0.10$ and $>0.05$, and the number of CSI
measurements within each bin, are given in Table \ref{tab:zz}.
\label{fig:zz}}
\end{figure*}

We tested our density-dependent completeness corrections in order to verify that the
effects of slit collisions do not systematically bias measurements of galaxy environment.
Using those galaxies with CSI redshifts to infer the projected local source densities, we
show the accuracy with which the intrinsic projected densities can be recovered in
Figure \ref{fig:dens}(right).
In the figure we plot the ratio of the inferred local source
densities for galaxies in several $i$-band magnitude bins to
true local densities of galaxies in those magnitude ranges. We plot the 16th, 50th, and 80th percentiles
as dashed, solid, and dashed lines, showing the error in inferred local source density as a function of the linear scale over which
the densities were computed. For comparison, 
the correlation of $R_{200}$ with $M_{vir}$
\citep[e.g.][]{carlberg1997,carlberg2001} implies circular diameters of $\sim$ 270 arcsec and
$\sim 120$ arcsec for halos of mass $M_{vir} = 5\times 10^{14}M_\odot$ and $M_{vir} = 5\times
10^{13}M_\odot$, respectively, at $z=1$. 
The systematic bias in the recovered local
source densities is less than 10\% at all magnitudes and scales larger than $\sim 2$ arcmin.

The distribution of inferred to total densities is asymmetric on the shortest
scales and at the faintest magnitudes; this is driven by small number statistics. Essentially,
when the inferred source density is zero, there may be two reasons:  the true source density may, in fact, be zero, or
there may be one
or two galaxies present that were simply missed by the slitmasks or rejected by our quality cuts.
The combination of slit collisions and the dependence of the completeness on
$i$-band flux imply an effective optical limit of $i\sim 24.5$ mag for environmental studies with CSI
for scales down to $\sim 2$ arcmin. 
The 1-$\sigma$ confidence intervals also indicate that the inferred local densities
of galaxies at a given magnitude have typical formal uncertainties of $\simlt 0.2$ dex. Thus, incompleteness, including the effects
of slit collisions, add $\simlt 0.2$ dex to the uncertainty in, for example, group masses when estimated from counting up galaxies
within radii $\simgt 2$ arcmin.

\subsection{Redshift Uncertainties}
\label{subsec:accuracy}

We have
three different methods at our disposal for evaluating the accuracy of CSI redshifts:  (1) directly
comparing our results with hundreds of optically-selected high-resolution spectroscopic redshifts from several sources;
(2) deriving uncertainties using the large number of objects with repeat measurements;
and (3) employing the \cite{quadri2010} pairwise velocity approach, in which galaxy groups and
large scale structures mean that a significant fraction of galaxy pairs are physically associated, effectively providing repeat redshift 
measurements of cosmic structures. In the following, we apply these three metrics to our CSI data to robustly test the reliability
of our redshifts (and their quoted uncertainties).

\subsubsection{Direct Comparisons with High-Resolution Spectroscopic Redshifts}
\label{sec:specz}

\begin{figure*}[t]
\centerline{
\includegraphics[width=7.0in]{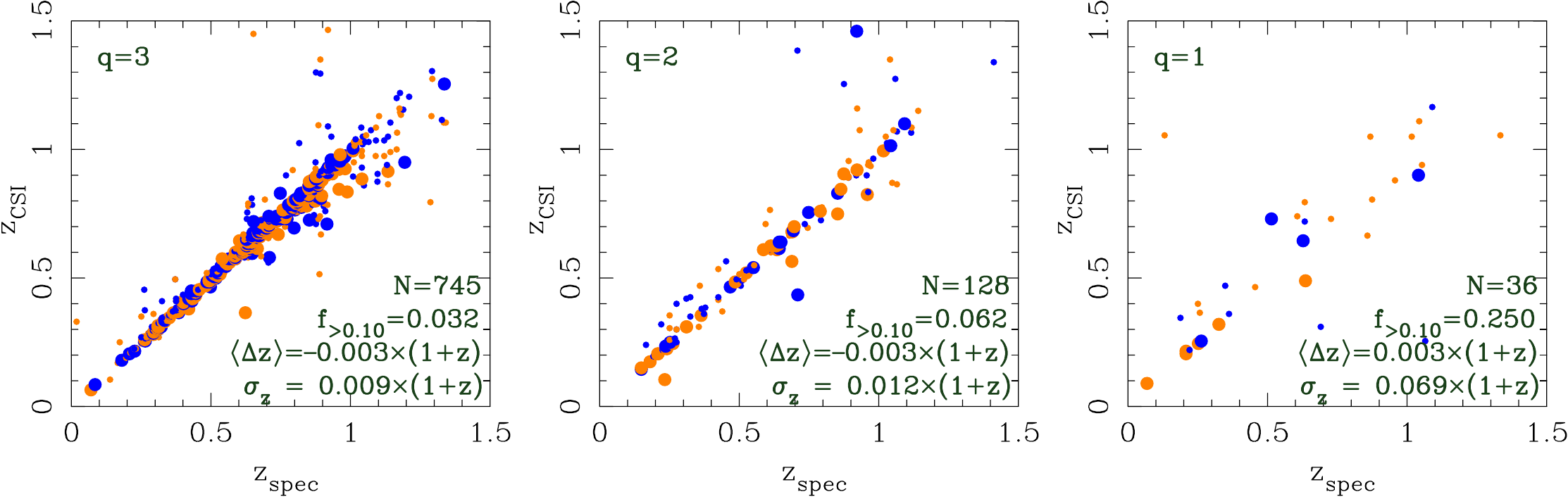}
}
\caption{CSI redshifts with $q=3,2,1$ plotted against redshifts in the VVDS
and in the UDS from \cite{cooper2013} with quality flags 3,4,5.
Larger symbols indicate higher quality redshift flags from the high resolution datasets.
\label{fig:zzq}}
\end{figure*}

Figure \ref{fig:zz} shows a direct comparison of previously published spectroscopic redshifts with those
derived by our SED fitting. Figure \ref{fig:zzq} breaks this comparison down by CSI data quality.
For the 227 high quality VVDS measurements that overlap our sample, 91\% of the redshifts agree to within $\delta z/(1+z)<0.02$, with 96\% within 
$\delta z/(1+z)<0.05$. The standard deviation for this subsample is $\sigma_z/(1+z)=0.007$
(using $\sigma=1.48\times$ MAD),
with a bias of $\langle \Delta z/(1+z)\rangle
= -0.004\pm0.0005$ (standard error of the mean).
For the 194 poorer quality VVDS measurements, we find a
scatter of $\sigma_z/(1+z)=0.015$, with 88\% agreeing to within $\delta
z/(1+z)< 0.05$, and a slightly larger bias of $\langle \Delta
z/(1+z)\rangle = -0.005\pm0.002$. This bias in the redshifts is roughly
equivalent to half a pixel, suggesting that systematic errors in our wavelength calibrations may remain at that level.
This offset is consistent with the small shift in wavelengths due to differences in the light path through the
instrument from the helium lamps compared to light from the sky, and future efforts to refine the data pipeline
will be devoted to reducing this bias.

\begin{deluxetable}{l l c c c c}
\tablecaption{The Comparison of CSI Redshifts
with Previously Published High Resolution Data
\label{tab:zz}}
\tablehead{
\multicolumn{2}{c}{Range} &
\colhead{$\sigma$} &
\colhead{$N$} &
\colhead{$N_{>0.10}$}&
\colhead{$N_{>0.05}$}
}
\startdata
\cutinhead{By Redshift}
\phn 0.125 & \phn 0.375 & 0.006 &     159  & \phn 2  & \phn 9 \\
\phn 0.375 & \phn 0.625 & 0.007 &     273  & \phn 3  & 17 \\
\phn 0.625 & \phn 0.875 & 0.012 &     286  & \phn 5  & 34 \\
\phn 0.875 & \phn 1.125 & 0.013 &     146  & \phn 6  & 23 \\
\phn 1.125 & \phn 1.375 & 0.027 & \phn 22  & \phn 7  & \phn 8 \\
\cutinhead{By $i$-band Magnitude}
18.000 & 19.300 & 0.006 & \phn 50  &\phn  1  & \phn 4 \\
19.300 & 20.600 & 0.007 &     238  &\phn  2  & 15 \\
20.600 & 21.900 & 0.008 &     315  &\phn  4  & 20 \\
21.900 & 23.200 & 0.015 &     225  &     10  & 36 \\
23.200 & 24.500 & 0.039 & \phn 44  &     11  & 20 \\
\cutinhead{By $[3.6]$ Magnitude}
18.000 & 18.600 & 0.006 & \phn 61  &\phn 0 & \phn 6 \\
18.600 & 19.200 & 0.007 &     202  &\phn 2 & 15 \\
19.200 & 19.800 & 0.009 &     200  &\phn 6 & 17 \\
19.800 & 20.400 & 0.010 &     215  &   11  & 31 \\
20.400 & 21.000 & 0.012 &     207  &   13  & 32 \\
\enddata
\end{deluxetable}


In Figure \ref{fig:zz} we show the $1 \sigma$ scatters in magnitude bins, also derived using $\sigma=1.48\times$ MAD.
Note that these values include a number of galaxies with low quality VVDS redshifts. For galaxies down to
$i=23$ mag, the scatter is $\sigma_z/(1+z)=0.009$. Fainter than $i=23$ mag,
the scatter rises to $\sigma_z/(1+z)=0.022$ (excluding those with $|\Delta z/(1+z)|>0.2$;
$\sigma_z/(1+z)=0.029$ when these outliers are included). For galaxies with
CSI redshifts between $0.9<z<1.2$, we find
$\sigma_z/(1+z)=0.018$, though the distribution is non-Gaussian and the
subsample contains a large fraction of objects with
low quality VVDS redshifts.
These standard deviations are reproduced in Table \ref{tab:zz} along with the numbers of outliers
with $|\delta z/(1+z)| > 0.10$ and $>0.05$, and the total number of CSI measurements within each
bin. Overall the median systematic bias in CSI redshifts is
$\langle \delta z/(1+z)\rangle = -0.004$.

The consistency between our redshifts and those made available by the VVDS 
confirms that our procedures are working well. However, the number of high-resolution spectroscopic
redshifts in the literature is too small
to accurately characterize our redshift uncertainties in bins of magnitude, color, or spectral
type using direct comparisons. Fortunately, we have two additional tests to help us probe the accuracy of our redshifts.

\begin{figure*}[h]
\centerline{
\includegraphics[width=7.0in]{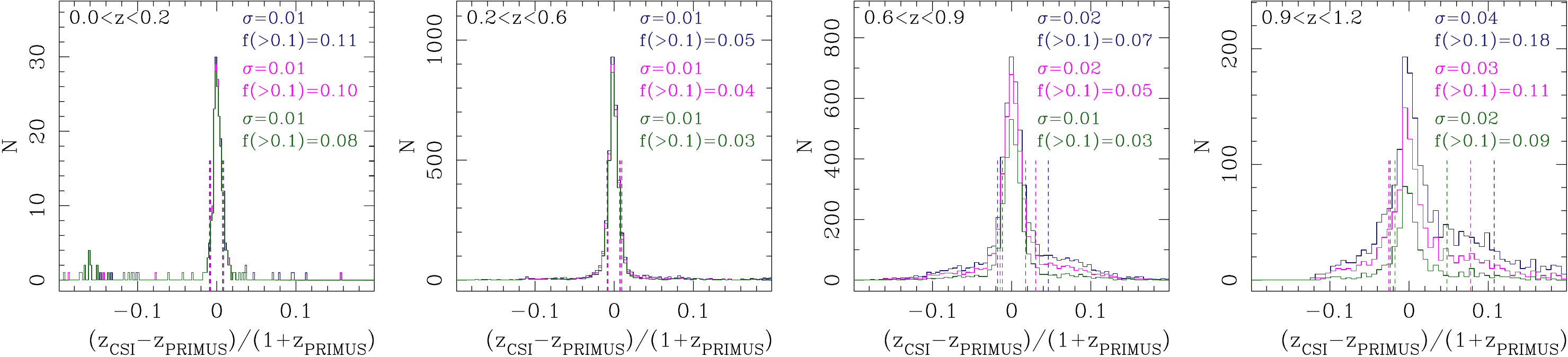}
}
\caption{A comparison of high quality CSI redshifts (with $q=3$) with those published by
PRIMUS \citep{cool2013} in four redshift ranges. The comparison is shown using
PRIMUS quality flags $2,3,4$ (blue), quality flags $3,4$ (violet), and quality flags $4$
(green). Overall agreement is excellent. Vertical lines show the 16th and 84th percentiles of
the distribution, highlighting the tails of the distributions, and for the higher redshift bins,
the dependence of the tails on PRIMUS quality flag. Above $z=0.6$ there is a pronounced tail of
galaxies with CSI redshifts significantly higher than those measured by PRIMUS. This tail arises
when galaxies with modestly strong [OII]3727\AA\ emission (and blended with the 4000\AA\ break)
are fit with templates having a Balmer break and low (or no) levels of [OII]3727\AA\ emission.
\label{fig:primus}}
\end{figure*}

\subsubsection{Direct Comparisons with Other Low-Resolution Redshifts}

The SWIRE XMM-LSS field was also observed by the PRIMUS collaboration \citep{coil2011,cool2013}
who have published 213,696 redshifts, with 70,430 in fields that overlap with the CSI data
presented here. In our ``high-quality'' CSI sample, 15,431 sources
also have prism redshifts made available publicly
by \cite{cool2013}. In Figure \ref{fig:primus} we show histograms of the fractional redshift
differences between CSI and PRIMUS for the 12,457 CSI measurements with $q=3$
in four increasing bins of redshift. \cite{cool2013} published these redshifts with
quality flags and we restrict the histograms to sets with PRIMUS flags $2,3,4$ (blue),
$3,4$ (violet), and just $4$ (green).

In each panel we show standard deviations of the distributions as
estimated from the MAD. We also show the fraction of CSI redshifts that differ from PRIMUS's
by more than 0.10 in $\delta z/(1+z)$.
Overall, the level of agreement is very good given the strikingly different observing strategies,
reduction techniques, and SED analysis. By itself this level of repeatability validates the
approach of low dispersion spectroscopy.
In the lowest redshift bin ($z<0.2$) there
is a tail to low CSI redshifts. At such low redshifts the $u$-band can be an
important redshift discriminant and some of this tail is due to our use of M-estimators that
reduce the weight of ``discrepant'' data points in fitting templates to the SEDs. When an
$r$-band point is discrepant, the effect is minimal due to the wavelength coverage of the
spectra. But at $u$ there are no spectroscopic data to anchor the fit, and the net effect is to
downweight the $u$ data (the issue is also relevant at $J$ and $K_s$). Future iterations of
improvements to the SED fitting are expected to mitigate this (small) problem.

At $z>0.6$, as shown in the last two panels, the distributions have sharp cores peaked at
zero but skew to positive redshift offsets compared to those from \cite{cool2013}. Based on our
experience, and the experiences of \cite{patel2010}, we suggest that this tail to higher CSI
redshifts is the result of (1) differences between our templates, and (2) the broader
wavelength baseline of the CSI SED fitting. Our theoretical templates have unrestricted emission
line components in contrast to empirical templates. With empirically derived templates,
the emission line measures are fixed by the set of galaxies (at lower redshift) that define the
template set. In PRIMUS, the mean redshift defining the templates is $z\sim 0.3$
\citep{cool2013}. Reproducing the potentially larger range of emission line luminosities (with
respect to the underlying stellar continuum properties) at earlier epochs may be
problematic.

\begin{figure*}[t]
\centerline{
\includegraphics[width=6.0in]{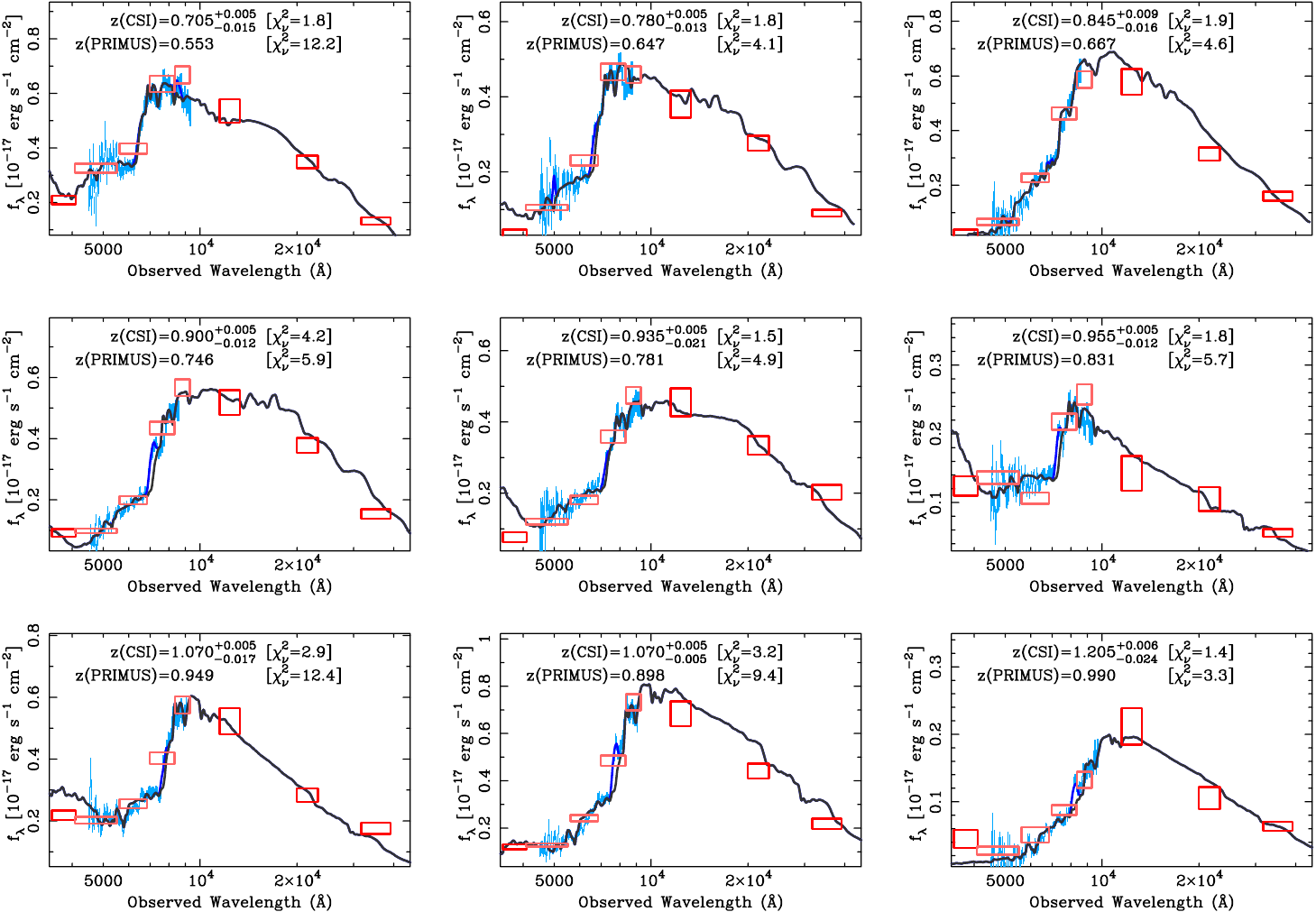}
}
\caption{Example galaxies from PRIMUS \citep{cool2013} where the CSI redshifts are higher, such that
$0.05 \Delta z/(1+z)<0.12$. Typically these galaxies have intermediate stellar populations, and the
combination of [OII]3727\AA\ line emission and the 4000\AA\ and Balmer breaks all
conspire with the low resolution to make fitting that region less robust. PRIMUS's SED
fitting did not include the wide wavelength baseline used by CSI, and thus could not provide the
same leverage on those stellar population components that define the Balmer and 4000\AA\ breaks.
Furthermore, the PRIMUS redshifts were derived using empirical templates, which do not
necessarily allow for the full range of emission line measures that may exist outside of the
redshift range that defined the templates.
\label{fig:primus2}}
\end{figure*}

\begin{figure*}
\centerline{
\includegraphics[width=6.5in]{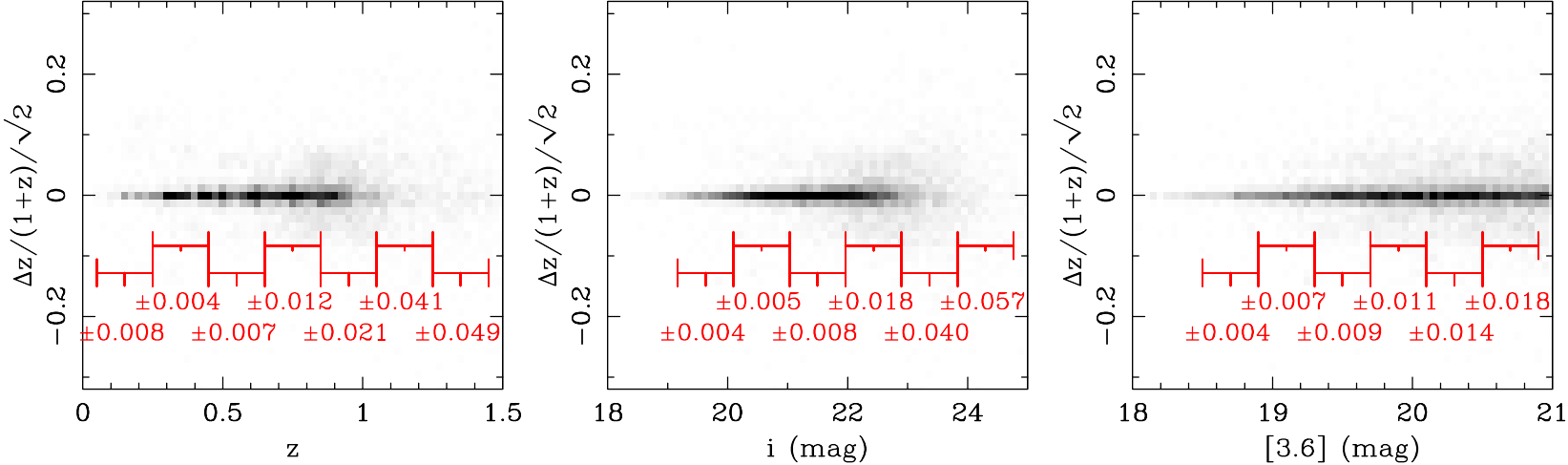}
}
\caption{Fractional redshift differences between repeat measurements, 
scaled by $\sqrt{2}$, as functions of redshift, $3.6\mu$m magnitude, and 
$i$-band magnitude for galaxies with $q\in \{2,3\}$.
For example, at $0.6<z<0.8$, the half-width of the 68\% confidence interval 
is $\sigma=0.009$. At $0.8<z<1.0$ we find  $\sigma=0.020$. As seen in the comparison
to VVDS, the width of the distribution is a stronger function of optical 
magnitude than near-IR magnitude, because our SED fitting is dominated by 
the IMACS spectroscopy.  We find $\sigma=0.033$ at 22.6 mag $<i<$ 23.6 mag, 
and $\sigma=0.053$ at 23.6 mag $< i <$ 24.6 mag. 
\label{fig:repeats_z}}
\end{figure*}

The allowed strength of [OII]3727\AA\ line emission is important when the resolution is
very low. Galaxies with weak (or no) [OII]3727\AA\ line emission and a
Balmer break indicative of intermediate age stars can sometimes be fit with templates that have [OII]3727\AA\
line emission blended with the 4000\AA\ break. This effect is similar to that seen in the low resolution spectroscopic
study of A stars by \cite{yu1926}, in which variable strengths of the lines in the Balmer series can shift the apparent
location of the break.
We show example SEDs in Figure \ref{fig:primus2} for galaxies at $z>0.6$ from the positive tail of the
distribution of redshift differences between CSI and PRIMUS, listing both redshifts and the
reduced $\chi^2$ we obtain for our best-fit templates at the CSI and PRIMUS redshifts.
Note that many of these galaxies have modest [OII]3727\AA\ and intermediate age stellar
populations. The wide baseline of broadband photometry used by CSI helps constrain those stellar
population components that define the 4000\AA\ and Balmer breaks. We also take
Figures 20 and 21 of \cite{cool2013} as confirmation that these tails to high CSI redshifts
are a reflection of these issues of resolution and feature strength. In those plots,
\cite{cool2013} shows that the PRIMUS redshift error distribution has a tail
to negative redshift errors for galaxies at $z\simgt 0.6$. We take the dependence of this tail on
PRIMUS's quality flag (and the dependence of the tail in Figure \ref{fig:primus} on the PRIMUS
quality flag) as suggestive that the tail is due to PRIMUS's systematic bias to low redshifts
(for galaxies of this particular spectral type).

\subsubsection{Uncertainties Derived from Repeat Measurements}

Because the data from individual slit masks were reduced independently, repeat measurements can be compared. These comprise
approximately 20\% of the measured redshifts, providing a sample large enough to be analyzed as a function of galaxy properties, 
such as magnitude, color, stellar mass, or redshift. Without prior constraints or measurements for the derived properties of these
objects we adopt the mean of multiple measurements when plotting or analyzing the distributions as functions of redshift. In 
Figure \ref{fig:repeats_z} we plot the 68\% fractional redshift differences against redshift, as well as $3.6\mu$m and 
$i$-band magnitudes. As a function of redshift the half-width of the 68\% interval is $0.009$ at $0.6<z<0.8$,
$0.020$ for $0.8<z<1.0$, and $0.034$ for $1.0<z<1.2$. 
As can be seen in the third panel, the width of the distribution is a stronger function of optical magnitude
than near-IR magnitude because our SED fitting is dominated by the IMACS spectroscopy. Currently, we find
$\sigma=0.033$ at 22.6 mag $<i<$ 23.6 mag, and $\sigma=0.053$ 
at 23.6 mag $< i <$ 24.6 mag. Planned refinements in the pipeline are expected to improve the fidelity of the extracted spectra for
such faint sources as our survey progresses. 

\begin{figure*}
\centerline{
\includegraphics[width=6.0in]{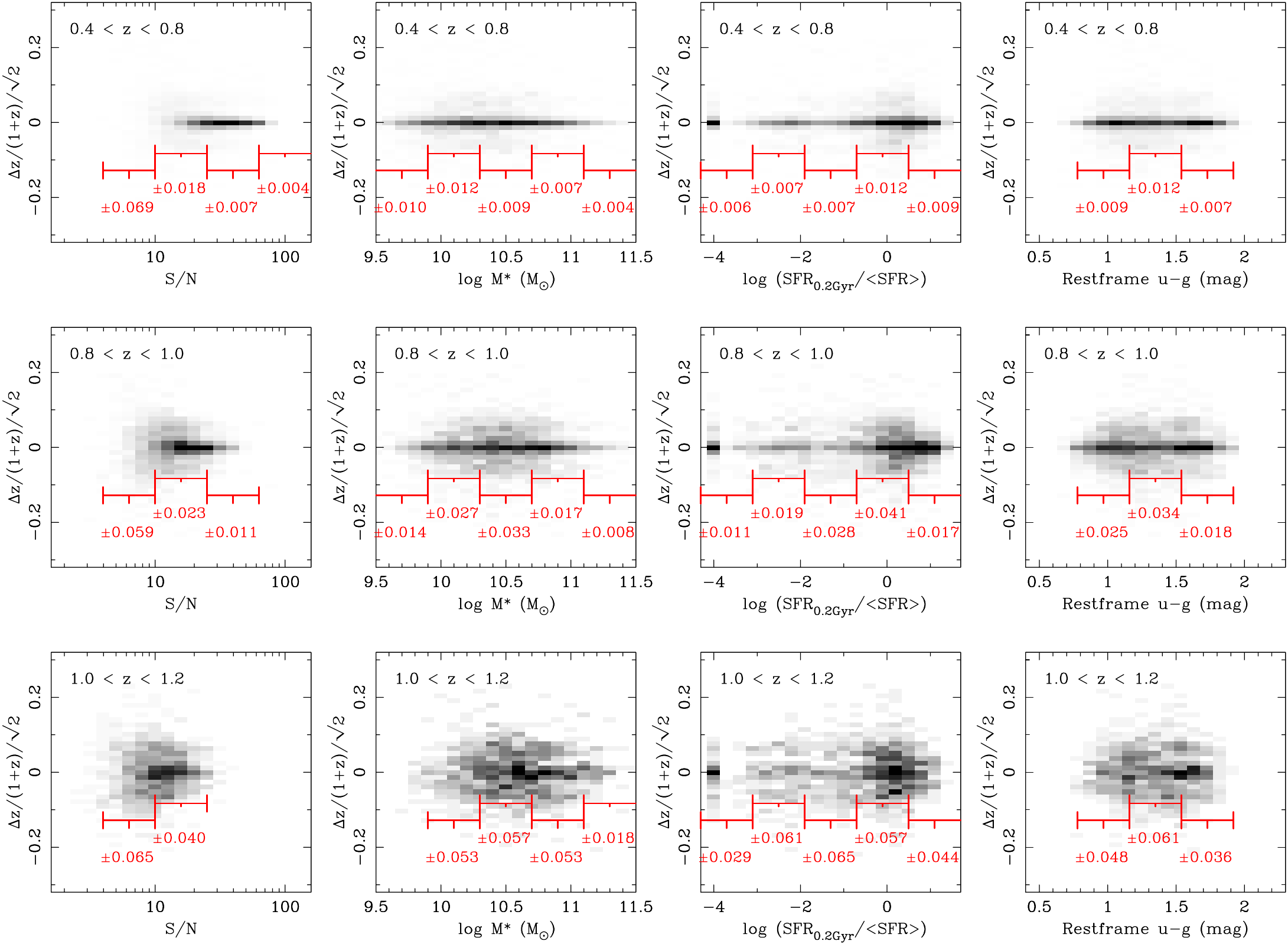}
}
\caption{Fractional redshift differences between repeat measurements, 
scaled by $\sqrt{2}$, as functions of $S/N$ ratio (75th percentile of the prism spectra), stellar mass, relative 
star formation activity, and restframe $u-g$ color for galaxies with $q\in\{2,3\}$. There is very little
dependence of the redshift errors on spectral type or color, unlike what is typically seen 
for photometric redshifts.
\label{fig:repeats_z2}}
\end{figure*}

The redshift uncertainties are dissected further in Figure \ref{fig:repeats_z2} by breaking up the
sample into three redshift ranges and plotting against $S/N$ ratios of the IMACS spectra, stellar
mass, star formation activity, and rest frame $u-g$ color. From these panels it is evident that the
redshift errors are not strongly correlated with spectral type or color, unlike the 
dependencies often seen in photometric redshifts.
There is a rather small, subtle dependence on spectral
type: quiescent galaxies (those with a negligible specific star formation rate and hence strong Balmer/4000\AA\ breaks) and strongly
star-forming galaxies have the greatest precision, while galaxies with modest or intermediate SFR
have degraded redshift precision.
Weak [OII]3727\,\AA\ emission at high redshift and observed at low resolution can appear blended with the
4000\AA\ break and produce spectra that are reasonably well fit by an SED with a Balmer break
and no [OII]3727\,\AA. More specifically, [OII]3727\,\AA\ is imperfectly correlated with the SFR
inferred from the stellar continuum, leading to an additional uncertainty in the redshifts at the
$\sim 0.01-0.02$ level in $\delta z/(1+z)$, depending on the $S/N$ ratio of the spectra, and even when not particularly
statistically significant.
This effect is substantially less pronounced for UDP data, where the higher resolution of the  disperser shifts the 
wavelength of such [OII]3727\AA\ confusion farther to the red. This power of the higher resolution
UDP is illustrated in the last SED of Figure \ref{fig:examples1}, in an old galaxy at $z=1.27$ with
[OII]3727\AA\ line emission \citep[likely a LINER, e.g.][]{lemaux2010}.
Again, this problem with low resolution spectroscopy is only important for ``green''
galaxies;  the bulk of the galaxy population, lying in the blue cloud and red sequence 
(with strong [OII]3727\,\AA\ and Balmer breaks respectively), is unaffected.

While the abundance of repeat spectroscopic measurements has allowed us to characterize the redshift
uncertainties, repeated spectra of any given object are fine-tuned with the same broadband photometry as described in Section \ref{subsec:fitting}.
For some galaxies this can in principle introduce artificially tight correlations between the repeat spectroscopic measurements.
Presumably, this issue becomes a serious problem only if the broad-band photometry is allowed to have greater weight in the
likelihood analysis. This never occurs in our procedures because the number of data points in the UDP and LDP
spectra far exceed the number of photometric bands. Even when the $S/N$ ratios of the spectra are low, binning these data over
the photometric bandpasses yield $S/N$ ratios exceeding those of the broadband photometry for faint galaxies.
However, since other factors (e.g., template selection) may produce artificially consistent repeat measurements, in the following section we
employ one more independent test of the redshift uncertainties.

\subsubsection{Uncertainties from Pairwise Velocity Distributions}

Because galaxies largely exist in close pairs, in groups, in clusters, and in even larger coherent structures, all of which have velocity 
widths smaller than the redshift errors we expect in our data and analysis, the distribution of redshift errors can be inferred statistically 
from the distribution of pairwise redshift differences in a given sample. This technique was first explored and described by
\cite{quadri2010} in an effort to derive the redshift uncertainties in photometric galaxy surveys, where a lack of spectroscopic follow-up
and limited templates can hamper one's ability to estimate uncertainties. Any methodology relying explicitly on clustering requires 
dense spatial sampling in order to extract a strong signal from the pairwise velocity histograms.  While photometric redshift surveys, 
by definition, result in the highest possible  spatial sampling, the spatial sampling of CSI is limited by slit collisions,  limiting
the number of galaxy pairs available to perform this analysis.

We divided our sample into subsets based on $i$-band magnitude and redshift and constructed pairwise velocity histograms. For the 
subsamples based on magnitude, we used those galaxies brighter than a given bin as a reference set, since those galaxies on average have 
more precise redshifts. While \cite{quadri2010} chose to randomize the positions in their dataset to correct for close pairs that arise only in projection, we opted to randomize the 
galaxy velocities since our slit-placement constraints lead to nontrivial spatial dependencies.
Care was taken to properly normalize the histograms derived from the randomized distributions, as the number
of galaxies in them, by
definition, is too large by the number of galaxies within coherent velocity structures.
Because this number cannot be precisely 
defined {\it a priori\/}, it must be deduced from the data --- from the wings of the histograms. We made
several realizations to reduce the Poisson noise imposed by the
randomization process.

\begin{figure}[htb]
\bigskip
\bigskip
\centerline{
\includegraphics[width=2.0in]{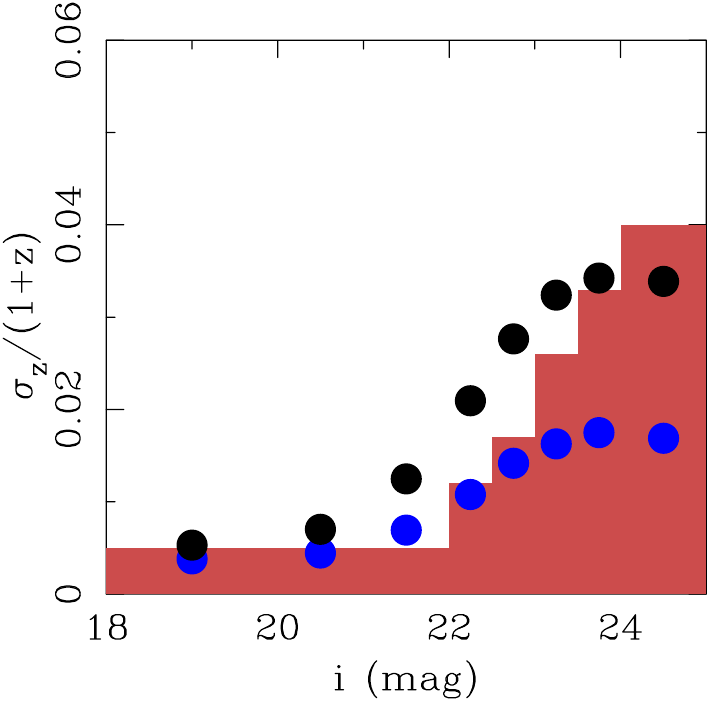}
}
\caption{A comparison of the redshift uncertainties derived from the pairwise velocity histogram method of
\cite{quadri2010} with the mean uncertainty derived from the redshift likelihood functions. The red shaded bars trace
the measured standard deviations from Gaussian fits to the pairwise velocity histograms as functions of
magnitude. Using blue circles we plot the mean uncertainties derived from the 68\% confidence limits
for the individual galaxies. The black circles are derived assuming that the 95\% confidence limits
better represent the true redshift distribution, such that for multimodal likelihood functions these
larger intervals better encompass any secondary (or tertiary) peaks. Overall, the
comparison is quite good, suggesting that the uncertainties from our redshift fitting procedures
are robust, with an increasing fraction of bimodal likelihood functions at faint magnitudes.
\label{fig:quadricompare}}
\end{figure}

\begin{figure*}[ht]
\centerline{
\includegraphics[width=5.0in]{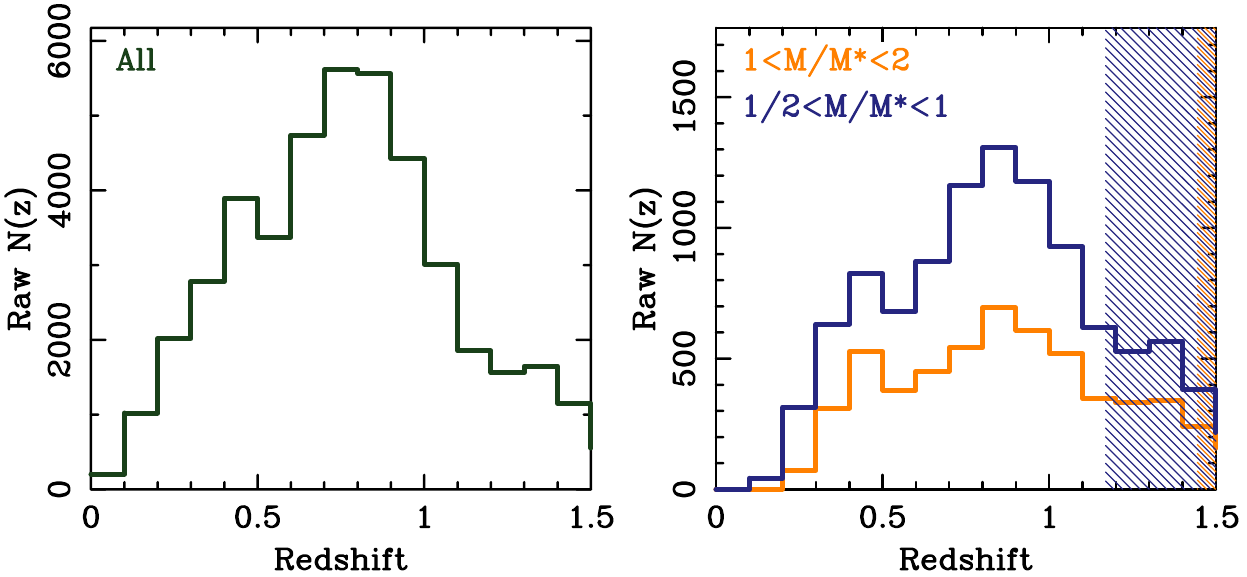}
}
\caption{
(top) Distributions of high quality CSI redshifts ($q\in\{2,3\}$).
(bottom) Distributions of high quality CSI redshifts for galaxies with stellar masses $1*<M/M*<2$ (orange)
and $1/2<M/M_*<1$ (blue). A large portion of the apparent decline in $N(z)$ above $z=0.8$ is due to an increase in
spectroscopic incompleteness (i.e. a declining success rate; see earlier discussion). The hatched regions mark
the redshifts where these subsamples are affected by selection biases, such that the optical and \emph{IRAC}
limits exclude old stellar populations of a given stellar mass. For galaxies $M>M_*$, CSI is statistically correctable
to $z\sim 1.45$. For galaxies $M>M_*/2$, CSI is correctable to $z=1.2$.
\label{fig:nz}}
\end{figure*}

The pairwise velocity histograms were fit by Gaussians and the resulting standard deviations as a function of magnitude are plotted in
Figure \ref{fig:quadricompare}. Overplotted in blue are the expectations using the 68\%
confidence limits from our redshift likelihood functions, which are consistent with the pairwise velocity histograms
down to $i \sim 23$ mag. We derived the black circles assuming that the 95\% confidence limits better represent the true
redshift error distribution. At faint optical magnitudes, the closer match of the black points with the red histogram
reflects a shift from unimodal to bimodal redshift likelihood functions, where the 95\% confidence limits probe both
peaks while the 68\% redshift confidence limits often do not fully encompass multiple peaks in bimodal (or trimodal)
likelihood functions. The consistency demonstrated by the \cite{quadri2010} method, along with comparisons to other spectroscopic redshifts, 
has reinforced the validity of the redshift uncertainties
derived from our generalized likelihood analysis down to faint magnitudes.
For faint sources, even when bimodality of the likelihood functions
becomes important, redshifts remain constrained on average to 0.05 or better in $\delta z/(1+z)$. This is consistent with the
relative numbers of galaxies with redshift errors $>0.10$ and errors $>0.05$ in $\delta z/(1+z)$ seen in 
\S \ref{sec:specz}.

The striking agreement between the different methods of estimating redshift errors
should be of general interest, since the \cite{quadri2010} method was in part devised to remedy the fact that 
photometric redshift surveys provide neither repeat measurements of a given galaxy nor high-resolution spectroscopy for
most targets. Since any spatially dense survey will 
provide repeat measurements of (relatively) cold cosmic structures, the \cite{quadri2010} methodology remains a
valuable technique for any spatially dense spectroscopic or photometric redshift survey --- any survey
where one does not have repeat observations with which to characterize uncertainties empirically.

Based on these three methods, we conclude that our random redshift uncertainties have been reliably characterized,
and also that systematic uncertainties appear to be minimal. In
particular we stress that our redshift errors are not strongly dependent on
relative star formation activity, spectral type, or color. As discussed
by \cite{quadri2012}, redshift errors that are strongly correlated with
spectral type will bias estimates of local galaxy density if care is not taken to ensure that velocity
windows (or line-of-sight linking lengths) encompass both red and blue galaxies with equal probability.
Given CSI's primary goal of characterizing galaxy properties as a function of environment, the relative insensitivity of
our redshift errors on spectral type is a crucial point.

\begin{figure*}[ht]
\centerline{
\includegraphics[width=5.0in]{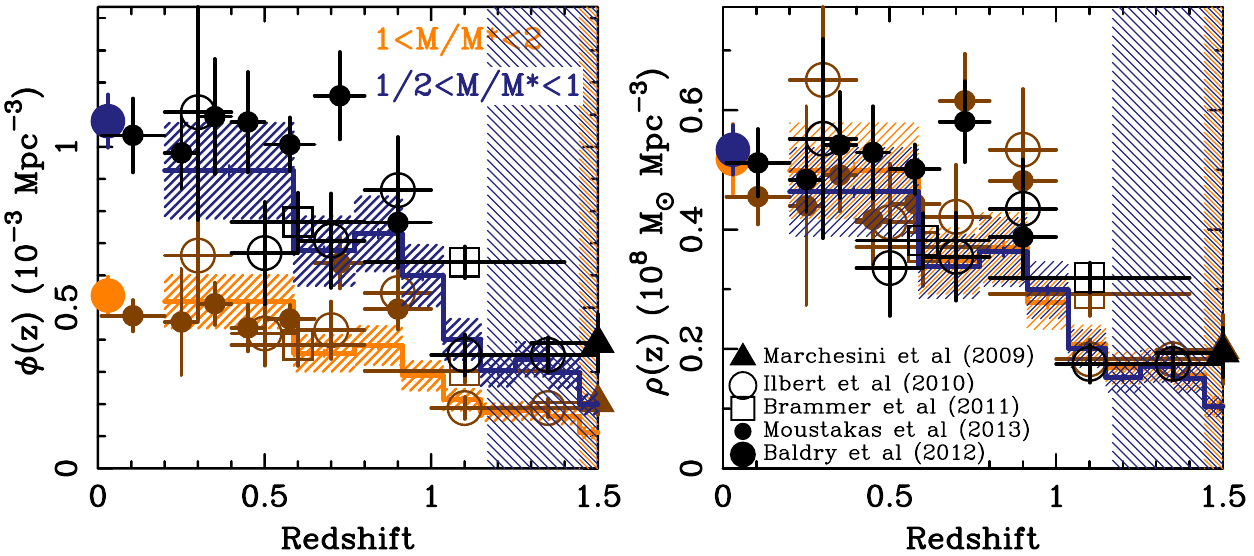}
}
\caption{
(top) Applying completeness corrections, we convert the $N(z)$ of Figure \ref{fig:nz} to the comoving
number density of galaxies with stellar masses $1<M/M*<2$ (orange thick line) and
$1/2<M/M_*<1$ (blue thick line).
The redshift bins are defined by a constant comoving volume in which cosmic variance is $\sim
15\%$ for red galaxies \citep[the hatched regions tracing the thick sold lines;][]{somerville2004}.
The large filled circles at $z=0.03$ are derived from GAMA
\citep{baldry2012}. The filled triangle at $z=1.5$ is derived from \cite{marchesini2009}. Open triangles are
derived from the mass functions of \cite{pozzetti2010}. The open squares derive from the mass functions of
\cite{brammer2011}. Open circles come from \cite{ilbert2010}. The galaxy number densities derived from PRIMUS
\citep{moustakas2013} are represented by the smaller filled circles. The agreement with these past results is
generally good with the exception of a large (2.5-$\sigma$) discrepancy between PRIMUS and CSI at $0.65<z<0.8$.
The blue hashed region beyond $z=1.2$ shows
approximately where the faint optical magnitude limit of the current analysis cuts the sample on the red
sequence (see Figure \ref{fig:masslimits}) at $M= M_*/2$. (bottom) Using the same completeness corrections
with the stellar mass estimates from the SED fitting, we compute the stellar mass density of the universe
contained in galaxies within the given two mass bins. The points are the same as in (top), and the agreement
between the stellar mass densities estimated here and past results is excellent.
\label{fig:phiz}}
\end{figure*}

\section{Discussion}
\label{sec:general}

We conclude our discussion of the survey by presenting a broad overview of the CSI sample. Using the sample directly with the
completeness functions derived earlier, we can estimate number densities of galaxies as a function of redshift and begin
looking at galaxy quiescence and star formation activity. As we do so, the CSI selection limits are highlighted in order to give the
reader a sense of the depth of the sample in luminosity, stellar mass, and redshift.

Figure \ref{fig:nz} (top) shows the distribution of high quality CSI redshifts ($q\in\{2,3\}$) in the CFHTLS XMM field. The decrease
in $N(z)$ above $z=0.8$ is due in part to the declining success rates and completeness at fainter optical magnitudes (as shown
earlier). Figure \ref{fig:nz} (bottom) dissects the sample into two stellar mass ranges, $1<M/M*<2$
(orange), and $1/2<M/M_*<1$ (blue). Our completeness functions derived earlier allow us to straightforwardly
correct the observed numbers of galaxies in CSI to accurate number densities down to magnitude limits of $[3.6]=21$ mag and $i=24.5$
mag. At stellar masses of $M=M*$ and $M=M*/2$, we estimate that the magnitude limits impose selection biases at $z\simgt
1.45$ and $z\simgt 1.2$, respectively; these limits are shown as hatched regions with the respective color.
Above these redshifts, the $i$-band and \emph{IRAC} flux limits
explicitly exclude galaxies from the selection and no statistically meaningful statements about the galaxy census can be made.
However, below these redshifts these subsamples remain statistically correctable. 

Knowing that our samples are relatively free from bias down to masses of $M>M*/2$ back to $z=1.2$, we
apply completeness corrections and convert our $N(z)$ figures to estimates of the comoving number densities
and stellar mass densities of galaxies as functions of redshift. The resulting CSI number densities are plotted using the thick
lines in Figure \ref{fig:phiz}. The number and mass densities are shown for two ranges of galaxy stellar mass,
orange for $1<M/M*<2$, and dark blue for $1/2<M/M_*<1$. Within these redshift bins the
estimated cosmic variance of $\sim 15\%$ dominates the uncertainties ($\sqrt{N}/N$ is small), and
this uncertainty is shown by the hatched regions surrounding the thick orange and blue lines.
The hatched regions shown at high redshift indicate redshifts where our faint magnitude limits
impinge on these subsamples and the \emph{IRAC} and $i$-band selections begin to exclude the oldest,
passive galaxies. Down to the limits of CSI, Figure \ref{fig:phiz} reinforces what previous deep surveys covering smaller
fields-of-view have found:  a factor of two increase in the number densities of normal galaxies since $z=1$
\citep[e.g.,][]{drory2009,marchesini2009,ilbert2010,moustakas2013}.

\begin{figure*}[ht]
\centerline{
\includegraphics[width=5.0in]{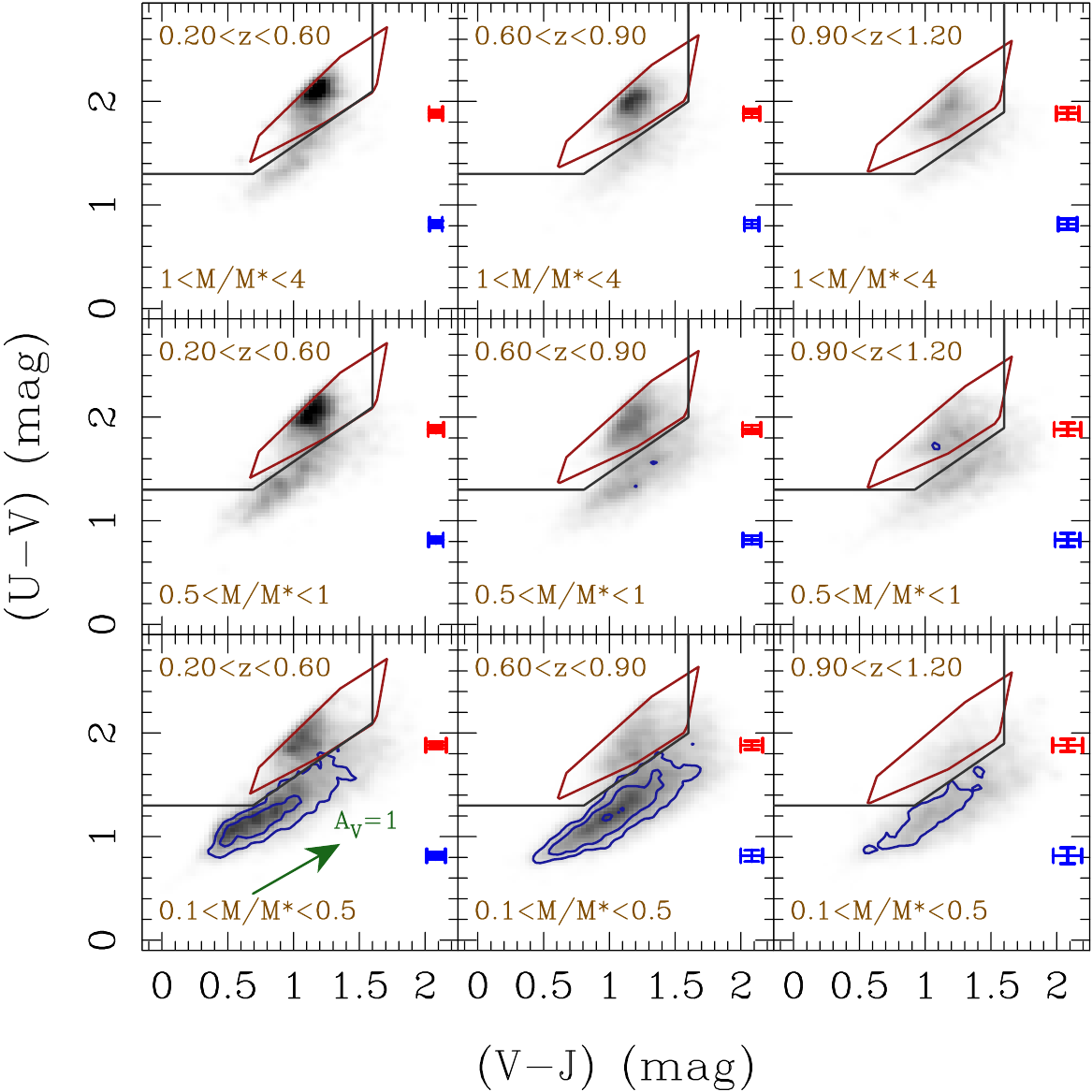}
}
\caption{Restframe $UVJ$ diagrams \citep[e.g.][]{williams2009} for galaxies in CSI in several bins of redshift and stellar
mass. The original \cite{williams2009} $UVJ$ selection boundaries for quiescent galaxies are shown in black, superimposed on grayscale maps of
the space density of galaxies at a given $U-V$ and $V-J$ color, mass, and redshift. The grayscales are fixed at a given mass scale
to a range defined by our lowest redshift bin, to help illustrate the change in number density at high redshift at fixed mass. A
green vector represents 1 mag of dust attenuation in the $V$-band for the \cite{calzetti2000} extinction law. Blue contours
highlight those galaxies with specific star formation rates $SSFR>2\times 10^{-10}$ yr$^{-1}$ (at $z=0.1$, evolving to $5.5\times
10^{-10}$ yr$^{-1}$ at $z=1$). These were derived using H$\alpha$ or [OII]3727\AA, depending on the redshift, corrected for dust
extinction using $A_V$ from the burst components of our SED fitting, and applying the conversion from \cite{kennicutt1998} or
\cite{kewley2004}. The red polygon shows the locus of galaxies with
$SSFR < 10^{-11}$ yr$^{-1}$ (at $z=0.1$, evolving to $2.8\times 10^{-10}$ yr$^{-1}$ at $z=1$) with a range of metallicities
from 1/10th to twice solar, and $0\le A_V\le 0.5$ mag of dust attenuation. We use these polygons to define a selection for
quiescent galaxies in the following figures. Note that for galaxies with stellar masses $M<M*/2$ at $z\simgt 1.2$, the flux limits of
CSI begin to select against quiescent galaxies, with these selection limits imposing on quiescent galaxies at lower masses at lower
redshifts (see, e.g., Figure \ref{fig:masslimits_lim}).
\label{fig:UVJ}}
\end{figure*}

\begin{figure}[ht]
\centerline{
\includegraphics[width=2.5in]{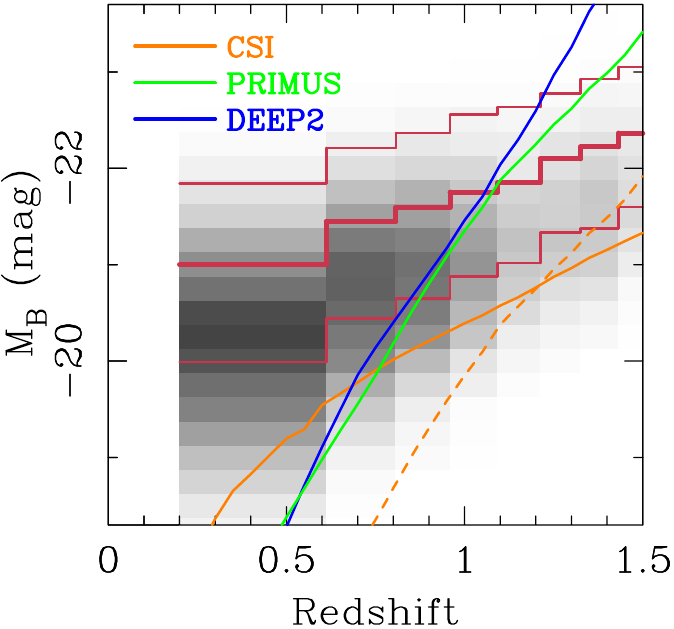}
}
\caption{The distribution of our $3.6\mu$m-selected galaxies in restframe $B$-band magnitude as a function of redshift.
The green and blue lines indicate the approximate loci for old stellar populations at the limits of the
PRIMUS and DEEP2 surveys. In orange, we show the CSI selection limit of $3.6\mu$m $=21$ mag (solid) and the effective
optical limit of $i=24.5$ mag (dashed).
\label{fig:bz}}
\end{figure}

While a detailed analysis of this evolution is beyond the scope of this data paper, the power of CSI's selection depth and field
size is illustrated by dissecting the sample in some basic ways.
Figure \ref{fig:UVJ} presents CSI bicolor diagrams for several ranges of stellar mass and redshift.
We have used our incompleteness functions and the $1/V_{max}$ method to compute the space densities of galaxies, represented
using the grayscales. By fixing the grayscale ranges to that derived in the low redshift bin, one can directly see the evolution
in number density at fixed mass that was shown in Figure \ref{fig:phiz}.

These $UVJ$ diagrams are directly comparable to those by other authors
\citep[e.g.][]{williams2009,whitaker2011}, with the red clump of quiescent galaxies and the blue star forming sequence seen
previously. The star forming sequence is especially prominent at low stellar masses. Using our H$\alpha$ and [OII]3727\AA\ emission
line luminosities to estimate rates of on-going star formation \citep[e.g.,][after correcting for dust
extinction]{kennicutt1998,kewley2004}, we identify those galaxies with high levels of star formation activity ($SSFR>2\times 10^{-10}$
yr$^{-1}$ at $z=0.1$, evolving to $5.5\times 10^{-10}$ yr$^{-1}$ at $z=1$) and show those 
with blue contours. These diagnostics reconfirm that the star-forming sequence in the $UVJ$ diagram remains an efficient means of
selecting star-forming galaxies.

The red polygon shows the locus of galaxies with $-1\le [Z/H]\le 0.3$ and $A_V\le 0.5$ and $SSFR < 10^{-11}$ yr$^{-1}$ at $z=0.1$
($2.8\times 10^{-10}$ yr$^{-1}$ at $z=1$). We can use this region to isolate galaxies with little, or no on-going star formation
(such loci can be generated for such regions in any color space for any filter set). At late times there appears to be a tail out of
the high-mass quiescent clump to blue colors along tracks such as those in Figure 4 of \cite{patel2011} for old populations with
very small recent star formation activity. But as Figure \ref{fig:phiz} already showed, there is a significant diminishing of
galaxies with stellar masses $M>M*/2$ at high redshift, and because quiescent galaxies dominate at those mass scales, 
the red clump, in particular, is quite degraded by $z\sim 1$.

Being able to track this evolution as a function of mass requires significant depth and we highlight this point
in Figure \ref{fig:bz}, showing the distribution of rest frame $B$-band absolute
magnitude as a function of redshift. The irregular redshift bins used here are defined to have constant volume over the CSI sample,
such that all bins have similar levels of cosmic variance \citep[$\sim 15\%$;][]{somerville2004}. The CSI $3.6\mu$m flux limit is
shown by the solid
orange lines, along with our effective optical limit of $i\approx 24.5$ mag as dashed orange lines. This $i$-band limit, where our
spectroscopic completeness falls below 1/3 of the maximum (of 50\% at $i=21$ mag, see Fig. \ref{fig:comp}), dominates over the IRAC
selection at $z>1.2$. For comparison, we plot the DEEP2 and PRIMUS magnitude limits of $r=24.1$ mag and $i=23.0$ mag in Figure
\ref{fig:bz} using the blue and green lines, respectively.
The red lines in Figure \ref{fig:bz} show the 10th, 50th, and 90th percentiles of the \emph{stellar mass density}
residing in quiescent galaxies. By $z=1$, DEEP2 or PRIMUS are already missing half the stellar mass in quiescent galaxies.

Surveys like DEEP2 or PRIMUS probe unbiased samples only down to the present day stellar mass of the Milky Way at $z=0.9$, and half
that at $z=0.75$. Although CSI's additional reach to $z=1$ for these latter galaxies Way may only extend 1 Gyr farther back in
cosmic time, the comoving volume covered within a constant area on the sky is nearly double that of PRIMUS. Such gains in
sensitivity highlight the importance of an infrared selection in crafting an unbiased picture of galaxies since $z=1.5$.
Furthermore, a methodology for measuring redshifts that does not depend crucially on the presence of emission lines at faint
magnitudes, ensures that the resulting samples are relatively free from bias against galaxies with low $M/L$ ratios at fixed mass.

\begin{figure}[h]
\centerline{
\includegraphics[width=2.75in]{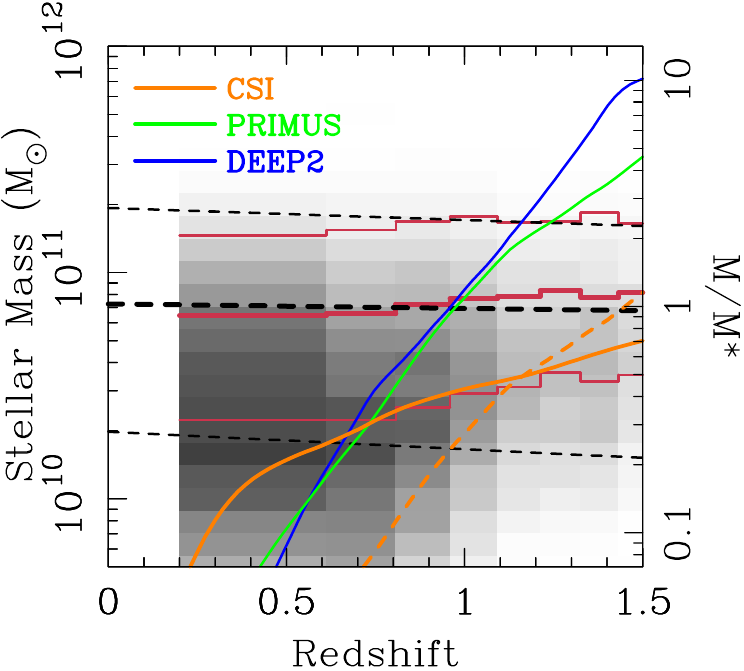}
}
\caption{The distribution of our $3.6\mu$m-selected galaxies in stellar mass, as functions of redshift
The green and blue lines indicate the approximate loci for old stellar populations at the limits of the
PRIMUS and DEEP2 surveys. In orange, we show the CSI selection limit of $3.6\mu$m $=21$ mag (solid) and the effective
optical limit of $i=24.5$ mag (dashed). The red contours trace the 10th, 50th, and 90th percentiles
of stellar mass density residing in quiescent galaxies. By carefully modeling the selection limits, we have computed corrections
to these percentiles as functions of redshift and have estimated their linear evolution with redshift. The resulting, underlying
evolution of the 10th, 50th, and 90th percentiles of the quiescent galaxy mass density are shown by the dashed black lines.
Our selection in the near-IR, depth in the optical, observing strategy, reductions, and analysis of the data enable CSI to reach
more than a magnitude fainter in the optical than DEEP2 or PRIMUS for passively evolving galaxies at $z=1$, enabling us to
accurately trace the bulk of the stellar mass of passive and active galaxies to $z\sim 1.4$ in an unbiased fashion.
\label{fig:mz}}
\end{figure}

\begin{figure*}[t]
\centerline{
\includegraphics[width=6.5in]{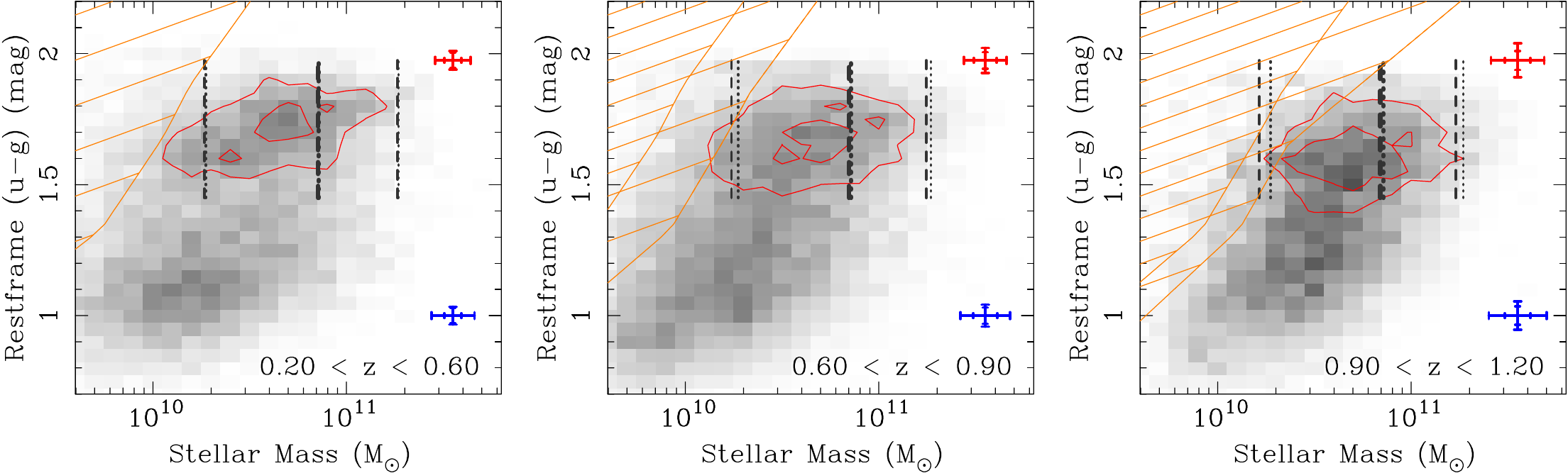}
}
\caption{Restframe color vs stellar mass for galaxies in three redshift bins in CSI, with the grayscale representing
the number densities of galaxies at a given color and stellar mass. The 10th, 50th, and 90th percentiles in stellar
mass density for quiescent galaxies at each epoch are shown by the vertical dashed lines. The vertical dotted lines show the
equivalent percentiles derived from the $0.2<z<0.4$ subsample, highlighting the very low level of evolution in the overall shape
of the quiescent galaxies mass function since $z\sim 1$. The orange hatched regions show the range of stellar mass and color where
the CSI flux limits impinge on the sample. The red contours show (arbitrary) levels of galaxy number density for galaxies selected
to be quiescent based on the $UVJ$ polygons of Figure \ref{fig:UVJ}. CSI is tracing the majority of stellar mass on the red
sequence, with minimal bias against passive galaxies back almost 9 Gyr in cosmic time.
\label{fig:ugm}}
\end{figure*}

To better illustrate this point, Figure \ref{fig:mz} shows a grayscale representation of the number densities of galaxies at a given
stellar mass and redshift. The 10th, 50th, and 90th percentiles of the stellar mass density residing in quiescent galaxies are shown
once again by the red lines. We model the selection differentially as a function of redshift, using the restframe colors of galaxies
in each redshift slice to determine which ones get excluded by the survey flux limits at progressively higher redshifts (described
in detail in a forthcoming CSI paper on the evolution of stellar mass and star formation), and derive linear fits to the evolution
of these percentiles of the quiescent galaxy population. This evolution, or lack therefore, is shown using the dashed black lines
in Figure \ref{fig:mz} and is also visualized in the color-mass diagrams of Figure \ref{fig:ugm}. The space densities of galaxies at
a given restframe $u-g$ color and stellar mass are shown in three redshift bins in CSI, with orange hatched regions marking the
projection of the CSI flux limits. The red contours trace the space densities of the quiescent galaxy population (see Figure
\ref{fig:UVJ}). The 10th, 50th, and 90th percentiles in the stellar mass density for quiescent galaxies are overlaid
using the vertical dashed lines, with the $0.2<z<0.4$ reference shown by the dotted lines.
Our selection in the near-IR, and depth in the optical, ensures that CSI can trace at least
90\% of the mass in quiescent galaxies back to early times. Each of these panels is normalized individually (unlike Figure
\ref{fig:UVJ}), such that the grayscales and contour levels reflect consistent levels relative to the number densities at their
respective epoch. The lack of any significant evolution in these percentiles (or in the red contours) indicates that those galaxies
that become quiescent do so without preference for a particular mass scale (at least for galaxies above $M\gg 10^{10}M_\odot$), even
as the number density of quiescent galaxies has more than doubled over the past 8.5 Gyr.

\section{Summary}
\label{sec:summary}

We have described the methodology behind, and initial data from, an ambitious new survey of approximately $2\times 10^5$ galaxies 
over a volume that encompasses the last 9 Gyr of cosmic history.  By selecting a sample at 3.6$\mu$m from
Spitzer-IRAC, the CSI survey is the most uniform to-date in terms of limiting mass as a function of redshift, and
will ultimately cover an unbiased volume comparable to SDSS but at $0.5<z<1.5$.  The power of
the \emph{IMACS} spectrograph with its low-resolution prisms allows us to survey large volumes efficiently and with sufficient
spectral resolution to detect large-scale structure and to measure emission lines from strongly star-forming galaxies and AGN.  CSI provides comparable redshift accuracies for red sequence and blue cloud galaxies; a significant advantage
compared to many broadband photometric---and even some spectroscopic---studies.  By combining our low-resolution 
spectrophotometry with extended broad-band photometry and sophisticated SED modeling, CSI bridges the
gap between surveys that are deep enough to probe galaxies below M$^*$ but are small in sample size and volume, and
those large-sample, large-volume surveys that do not reach typical, Milky Way-like galaxies that are the main event in
the history of cosmic evolution. In forthcoming papers we will update with more detail what we have
learned from the first batch of CSI survey data.

One of the highest priority goals will be to take advantage of this large volume to simultaneously measure
the evolution of the group mass function \citep[e.g.][]{williams2012},
and the dependence of star formation on environment \citep[e.g.][]{kelson2014}.
In doing so CSI will uncover the extent to which the growth in the number density of passive galaxies
over the past two-thirds of the lifetime of the universe can be attributed to group-related processes such as, for example, galaxy-galaxy
interactions (leading to differential number density evolution between group and field galaxies),
and/or a decrease in the available fuel for on-going star formation (leaving galaxy number densities
unchanged). Given that most galaxies in virialized or massive groups (or both) are ``red and dead'' even at $z\sim 1$ 
\citep{kelson2014}, the mass selection, high spectral coverage, and color-insensitive
redshift accuracy of CSI will all be critical elements in characterizing relatively poor groups at
high redshift. Subsequent CSI analyses will also focus on, for example, AGN and any
connections to group and/or host galaxy properties, on differences between centrals and
satellites \citep[e.g.][]{berlind2002,weinmann2006,vdb2007}, and on merger rates \citep[e.g.][]{williams2011} as functions of environment.

\section{acknowledgments}

D.D.K. expresses his appreciation to his
co-investigators for their patience. The whole team also appreciates the enormous contributions of the
Carnegie Institution for Science to the project, from the new disperser in IMACS, to the computing
cluster in the basement, to the generosity of the
Time Assignment Committee. Furthermore, the comments from the anonymous referee are acknowledged;
they were enormously helpful, in particular, in prompting us to improve the discussions of
incompleteness, and inspiring us to rethink how to quantify data quality.
We are grateful to NOAO for its contributions and to the astronomical community
for its awarding of survey status to CSI in 2009. R.J.W. also gratefully acknowledges support from NSF
Grant AST-0707417. Based, in part, on observations obtained with MegaPrime/MegaCam, a joint project
of CFHT and CEA/DAPNIA, at the Canada-France-Hawaii Telescope (CFHT) which is operated by the
National Research Council (NRC) of Canada, the Institut National des Science de l'Univers of the
Centre National de la Recherche Scientifique (CNRS) of France, and the University of Hawaii. This
work is based in part on data products produced at TERAPIX and the Canadian Astronomy Data Centre as
part of the Canada-France-Hawaii Telescope Legacy Survey, a collaborative project of NRC and CNRS.
This publication makes use of data products from the Two Micron All Sky Survey, which is a joint
project of the University of Massachusetts and the Infrared Processing and Analysis
Center/California Institute of Technology, funded by the National Aeronautics and Space
Administration and the National Science Foundation

\end{document}